\documentclass[aps, prd,twocolumn,superscriptaddress,nofootinbib,preprintnumbers]{revtex4-1}
\usepackage[utf8]{inputenc}
\usepackage[english]{babel}
\usepackage{natbib}
\def\apjl{ApJ}%
\def\apjs{ApJS}%
\def\ao{Appl.~Opt.}%
\def\aap{A\&A}%
\usepackage{graphicx}
\usepackage{amsmath,amssymb}
\usepackage{hyperref}
\usepackage{braket}
\usepackage{float}
\usepackage[dvipsnames]{xcolor}
\usepackage{soul}
\usepackage{rotating}
\usepackage{multirow}
\usepackage{mathtools}
\usepackage{makecell}

\usepackage[caption = false]{subfig}
\usepackage{txfonts}
\usepackage[margin=0.75in]{geometry}
\usepackage{tabularx}
\let\oldequation\equation
\let\oldendequation\endequation
\renewenvironment{equation}
  {\linenomathNonumbers\oldequation}
  {\oldendequation\endlinenomath}

\usepackage{hyperref}
\usepackage[T1]{fontenc}
\usepackage{ae,aecompl}
\usepackage{graphicx}	% Including figure files
\usepackage{amsmath}	% Advanced maths commands
\usepackage{adjustbox}
\usepackage{multirow}
\usepackage{newtxmath}  % For Indicator function. See https://tex.stackexchange.com/questions/26637/, answer by xovee.
\usepackage{booktabs}
\newcommand{\planck}{\textsc{Planck}}

\newcommand{\GalaxyMatterBias}{\ensuremath{b\,}}

\defcitealias{paperA}{paper I}
\newcommand{\paperA}{\citetalias{paperA}}

\newcommand{\avg}[1]{\langle #1 \rangle}

\newcommand{\imagunit}{\ensuremath{\textrm{i}\mkern1mu}}

\DeclareMathOperator{\PH}{PH}
\DeclareMathOperator{\Avg}{Avg}
\DeclareMathOperator{\WPHG}{WPHG}
\DeclareMathOperator{\ST}{ST}

\definecolor{purple}{RGB}{150,0,200}

\usepackage{lineno}
\begin{document}		%SWITCH TO PRD
%\label{firstpage} % SWITCH TO MNRAS
%\pagerange{\pageref{firstpage}--\pageref{lastpage}}% 

\title[DES Y3 moments, wavelet phase harmonics, and scattering transform]{Dark Energy Survey Year 3 results: simulation-based cosmological inference with wavelet harmonics, scattering transforms, and moments of weak lensing mass maps II. Cosmological results}

\label{firstpage}

%\makeatletter
%\def \blfootnote{\xdef\@thefnmark{}\@footnotetext}
%\makeatother

%\input{authors_MNRAS.tex}		%SWITCH TO MNRAS
%\input{authors_PRD.tex}		%SWITCH TO PRD

%\date{\today}

%\begin{document}     %SWITCH TO MNRAS
%\maketitle        	 %SWITCH TO MNRAS

% Abstract of the paper
\begin{abstract}
We present a simulation-based cosmological analysis using a combination of Gaussian and non-Gaussian statistics of the weak lensing mass (convergence) maps from the first three years (Y3) of the Dark Energy Survey (DES). We implement: 1) second and third moments; 2) wavelet phase harmonics; 3) the scattering transform. Our analysis is fully based on simulations, spans a space of seven $\nu w$CDM cosmological parameters, and forward models the most relevant sources of systematics inherent in the data: masks, noise variations, clustering of the sources, intrinsic alignments, and shear and redshift calibration. We implement a neural network compression of the summary statistics, and we estimate the parameter posteriors using a simulation-based inference approach. Including and combining different non-Gaussian statistics is a powerful tool that strongly improves constraints over Gaussian statistics (in our case, the second moments); in particular, the Figure of Merit $\textrm{FoM}(S_8, \Omega_{\textrm{m}})$ is improved by 70 percent ($\Lambda$CDM) and 90 percent ($w$CDM). When all the summary statistics are combined, we achieve a 2 percent constraint on the amplitude of fluctuations parameter $S_8 \equiv \sigma_8 (\Omega_{\textrm{m}}/0.3)^{0.5}$, obtaining $S_8 = 0.794 \pm 0.017$ ($\Lambda$CDM) and $S_8 = 0.817 \pm 0.021$ ($w$CDM), and a $\sim$10 percent constraint on $\Omega_{\rm m}$, obtaining $\Omega_{\rm m} = 0.259 \pm 0.025$ ($\Lambda$CDM) and $\Omega_{\rm m} = 0.273 \pm0.029$ ($w$CDM). In the context of the $w$CDM scenario, these statistics also strengthen the constraints on the parameter $w$, obtaining $w<-0.72$. The constraints from different statistics are shown to be internally consistent (with a $p$-value>0.1 for all combinations of statistics examined). We compare our results to other weak lensing results from the DES Y3 data, finding good consistency; we also compare with results from external datasets, such as \planck{} constraints from the Cosmic Microwave Background, finding statistical agreement, with discrepancies no greater than $<2.2\sigma$.

%Beyond 2-point statistics contain additional information on both cosmological and astrophysical and observational (systematics) parameters. In this data paper we provide and end-to-end analysis of a set of non-Gaussian lensing statistics using detailed Year 3 data from the Dark Energy Survey. We implement directional wavelets (known as wavelet phase harmonics, WPH), scattering transforms and moments. These transforms capture information at all orders while being either linear or quadratic in the observed field, and thus more robust to noise. With scale cuts and implementation of all relevant systematics, we find that either of these statistics are significantly more constraining that the second and third order moments of the convergence field. Compared to the standard 2-point analysis, the gain in the dark energy figure of merit is ?? percent. We also show the improvements in some of the key systematics parameters. Finally, we implement a carefully validated Likelihood Free Inference approach to parameter estimation.  In a companion paper we present the methodology and its validation on mock catalogues. 
\end{abstract}
\author{M.~Gatti}\email{marcogatti29@gmail.com}
\affiliation{Department of Physics and Astronomy, University of Pennsylvania, Philadelphia, PA 19104, USA}
\author{G.~Campailla}
\affiliation{Department of Physics, University of Genova and INFN, Via Dodecaneso 33, 16146, Genova, Italy}
\author{N. Jeffrey}
\affiliation{Department of Physics \& Astronomy, University College London, Gower Street, London, WC1E 6BT, UK}
\author{L. Whiteway}
\affiliation{Department of Physics \& Astronomy, University College London, Gower Street, London, WC1E 6BT, UK}
\author{A. Porredon}
\affiliation{Ruhr University Bochum, Faculty of Physics and Astronomy, Astronomical Institute (AIRUB), German Centre for Cosmological Lensing, 44780 Bochum, Germany}
\author{J. Prat}
\affiliation{Kavli Institute for Cosmological Physics, University of Chicago, Chicago, IL 60637, USA}
\author{J. Williamson}
\affiliation{Department of Physics \& Astronomy, University College London, Gower Street, London, WC1E 6BT, UK}
\author{M.~Raveri}
\affiliation{Department of Physics, University of Genova and INFN, Via Dodecaneso 33, 16146, Genova, Italy}
\author{B. Jain}
\affiliation{Department of Physics and Astronomy, University of Pennsylvania, Philadelphia, PA 19104, USA}
\author{V. Ajani}
\affiliation{Department of Physics, ETH Zurich, Wolfgang-Pauli-Strasse 16, CH-8093 Zurich, Switzerland}
\affiliation{Department of Astronomy and Astrophysics, University of Chicago, Chicago, IL 60637, USA}
% ...
\vspace{0.2cm}

\preprint{DES-2023-0820}
\preprint{FERMILAB-PUB-24-0255-PPD}

%\blfootnote{$^{\star}$ E-mail: marcogatti29@gmail.com}
%\blfootnote{Affiliations are listed at the end of the paper.}
%\setcounter{footnote}{1}

\maketitle

%\blfootnote{Affiliations are listed at the end of the paper.}
\setcounter{footnote}{1}

\section{Introduction}

%\NJ{Is it ok to make sure we say Gower Street simulation suite?}

%\NJ{LFI or SBI? SBI in the title, but LFI in the text}

%\NJ{Goodness of fit is on the compressed statistics right? If so, this should be made clear, because typically the goodness-of-fit in compressed space is much much better (even when there are systematics)}

Weak gravitational lensing serves as an efficient technique for investigating the large-scale structure and matter distribution throughout the Universe. This method enables us to infer the distribution of matter lying in the foreground by examining the minor distortions evident in the shapes of background galaxies. Such effects of weak gravitational lensing provide a window into the history of the Universe's expansion and its geometric properties, acting as tools for understanding the underlying structure of the cosmos.

Conventional methods of cosmological analysis employ Gaussian statistics, using tools such as two-point correlation functions or power spectra to analyze the lensing signals. The most up-to-date analyses of this type are from the Dark Energy Survey (DES,~\citealt{y3-cosmicshear1,y3-cosmicshear2}), the Kilo-Degree Survey (KiDS, ~\citealt{Asgari_2021,li2023kids}), and Hyper Suprime-Cam (HSC, ~\citealt{HSC1,HSC2}).  A growing number of studies, however, have investigated and highlighted the role of non-Gaussian statistics in improving cosmological constraints, as the lensing observables carry information beyond that probed by standard Gaussian statistics. Examples of non-Gaussian statistics investigated include higher-order moments \citep{VanWaerbeke2013,Petri2015,Vicinanza2016,Chang2018,Vicinanza2018,Peel2018, G20,moments2021,Porth2021},  peak counts \citep{Dietrich2010, Kratochvil2010, Liu2015, Kacprzak2016, Martinet2018, Peel2018, Shan2018, ajani_peaks,Zuercher2021,HD2022,Zuercher2022}, one-point probability distributions \citep{Barthelemy2020,Boyle2021,Thiele2020}, Minkowski functionals \citep{Kratochvil2012,Petri2015,Vicinanza2019,Parroni2020},  Betti numbers \citep{Feldbrugge2019,Parroni2021}, persistent homology \citep{Heydenreich2021,Heydenreich2022}, scattering transform coefficients \citep{Cheng2020,Valogiannis2022a,Valogiannis2022,Cheng2024}, wavelet phase harmonic moments \citep{Allys2020}, kNN and CDFs \citep{Anbajagane2023,Banerjee2023},  map-level
inference \citep{Porqueres2022,Boruah2022}, and machine-learning methods \citep{Ribli2018, Fluri2018,Fluri2019,jeffrey_lfi,Lu2023}. 

This work uses a comprehensive simulation-based inference approach to analyze the weak lensing data from the first three years of the Dark Energy Survey (DES Y3). Our analysis concentrates on three distinct statistics: second and third-order moments, wavelet phase harmonic moments, and the scattering transform. While the second and third moments have been previously applied to DES Y3 data \citep{moments2021}, their implementation was based on a theoretical framework for the modelling of the moments. In contrast, our current approach is driven by simulations; i.e., we use simulations to model our observables. Moreover, this research is the first application of wavelet phase harmonic moments to weak lensing data, and one of the first applications of scattering transform (see \citep{Cheng2024}). These two statistics have become prominent for their similarities to convolutional neural networks (CNNs), designed to extract information from fields in a manner comparable to CNNs \citep{Mallat2020}, but without the requirement for training. In \cite{paperA} (hereafter \paperA{}), we showed that these non-Gaussian statistics provide additional information beyond that of third moments, indicating their potential utility.

We use the \textit{Gower Street simulations} (\citealt{jeffrey2024dark}), a suite of N-body simulations spanning a seven-dimensional $w$CDM parameter space. We incorporate key sources of systematic uncertainties in our forward model such as photometric redshift uncertainties, shear calibration errors, intrinsic alignments, and the effects of source clustering \citep{sourceclustering}. We implement an efficient neural network compression of our summary statistics, and we estimate the parameter posteriors via neural density estimation of the likelihood surface, using flexible neural networks that do not imposing restrictive assumptions about the likelihood or data model. This paper builds upon a companion work (\paperA{}), in which the methodology was validated using simulations and a series of systematic observational tests were conducted to verify the robustness of the analysis. This work applies the validated methodology to the DES Y3 data to yield cosmological constraints; we also compare these findings with results from other DES Y3 analyses and from external datasets.

A companion DES analysis (\citealt{jeffrey2024dark}) uses the same simulation-based inference pipeline as described in this paper, but using convolutional neural networks (map-level inference), peak counts, and power spectra to infer cosmology. Their results are consistent with the results presented in this work. All of these are analyses of DES Y3 data, pending the final full DES Year 6 data.

This paper is organized as follows: Section 2 summarizes the survey data as well as the simulations used for our model predictions and validation. Section 3 briefly describes the summary statistics used in this paper and the data compression. Section 4 briefly describes our simulation-based inference pipeline for parameter inference. Section 5 presents the pre-unblinding tests, and Section 6 discusses our main results, together with comparisons against other DES results and external probes.

\section{Data and simulations}

\subsection{DES Y3 weak lensing catalogue and weak lensing mass maps}

We use the DES Y3 weak lensing catalogue for our analysis \citep*{y3-shapecatalog}. This extensive catalogue comprises 100,204,026 galaxies, offering a weighted effective galaxy number density of 5.59 galaxies per square arcminute across a 4139 square degree area. The catalogue is based on the \textsc{METACALIBRATION} algorithm \citep{HuffMcal2017, SheldonMcal2017}, which calculates self-calibrated shear measurements from the multi-band, noisy images of observed objects. To correct for any residual calibration issues such as multiplicative shear bias, we incorporate adjustments based on sophisticated image simulations \citep{y3-imagesims}. The catalogue includes a per-galaxy inverse variance weight, which enhances the signal-to-noise ratio of our measurements. The galaxies are partitioned into four tomographic bins of roughly equal number density, as described in \cite{y3-sompz}.

The redshift distributions for these bins are derived using the SOMPZ method \citep*{y3-sompz}, augmented by clustering-based redshift information \citep*{y3-sourcewz} and a correction for the redshift-dependent influences of blending  \citep{y3-imagesims}. We produce weak lensing mass maps for each tomographic bin using a full-sky extension of the Kaiser and Squires algorithm \cite{KaiserSquires}, \citep*{y3-massmapping}. The maps are pixelized using \textsc{HEALPIX} \citep{2005ApJ...622..759G} at a resolution of \textsc{NSIDE} = 512; this yields a pixel size of approximately 6.9 arcminutes.

\subsection{Simulations}

\begin{table}
%\tiny
\caption {Model parameters (first column), their distribution in the Gower Street sims and in the mock catalogues derived from these sims (second column), and the prior used in the cosmological analysis (third column, shown only where different from the second column). The analysis prior can differ from the distribution of the samples as long as these parameters have been explicitly used during the training of the Neural Density Estimators (NDEs) when learning the likelihood surface; see \S \ref{sect:LFI} for more details. $^*$ In our simulation runs we usually excluded values of $w$ less than $-1$, but 64 simulations were run without this constraint. These were still used to train our NDEs, although we applied a strict prior of $w > -1$ for the analysis.} %\LW{Do we need to define $\mathcal{N}$ and $\mathcal{U}$? I'm happy not to.}}
\centering
\begin{tabular}{c c c}
\toprule
\textbf{Parameter} & \textbf{Mocks parameters} & \textbf{Analysis prior} \\
 & \textbf{distribution} & \\
\midrule
$\Omega_{\textrm{m}}$ & mixed active-learning  & $ \mathcal{U}(0.15,0.52)$ \\ & in $\mathcal{U}(0.15,0.52)$   \\ 
$S_8$ & mixed active-learning  & $\mathcal{U}(0.5,1.0)$ \\& in $\mathcal{U}(0.5,1.0)$ \\
\midrule
$w$ & $\mathcal{N}(-1,\frac{1}{3})$ for $-1<w<-\frac{1}{3}$ & $\mathcal{U}(-1,-\frac{1}{3})$ \\
 & $0$ ${\textrm{else}}^{*}$  \\
$n_s$ & $\mathcal{N}(0.9649, 0.0063)$   \\
$h$ & $\mathcal{N}(0.7022, 0.0245)$ \\
$\Omega_{\textrm{b}}h^2 $&  $ \mathcal{N}(0.02237, 0.00015)$\\
$m_{\nu} $& $ \exp(\mathcal{U}[\log(0.06), \log(0.14)])$ \\
\midrule
$A_{IA}$ & $\mathcal{U}[-3, 3]$  \\
$\eta_{IA}$ & $\mathcal{U}[-5, 5]$ \\
$m_{1}$ & $\mathcal{N}(-0.0063,0.0091)$ \\
$m_{2}$ & $\mathcal{N}( -0.0198,0.0078)$ \\
$m_{3}$ & $\mathcal{N}( -0.0241,0.0076)$ \\
$m_{4}$ & $\mathcal{N}(-0.0369, 0.0076)$ \\
$\bar{n}_i(z)$ & $p_{\textsc{HyperRank}}(\bar{n}_i(z) | x_{\textrm{phot}})$  \\ 
\bottomrule
\end{tabular}
\label{parameter}
\end{table}

Here is a brief summary of the simulations used in this work; see \paperA{} for details. 

The Gower Street simulation suite \citep{jeffrey2024dark} is key to our inference process. The mocks created from the Gower Street suite serve two purposes: they are used both for compressing the summary statistics and for performing the cosmological inference.  The suite consists of 791 gravity-only full-sky N-body simulations, produced using the \textsc{PKDGRAV3} code \citep{potter2017pkdgrav3}. The simulations span a seven-dimensional parameter space in $w$CDM ($\Omega_{\textrm{m}}$, $\sigma_8$, $n_s$, $h \equiv h_{100}$, $\Omega_{\textrm{b}}$, $w$, $m_{\nu}$); see Table \ref{parameter}. This table uses the parameter $S_8 \equiv \sigma_8(\Omega_{\textrm{m}}/0.3)^{0.5}$.

Each full-sky simulation can be divided into four non-overlapping DES sky footprints, yielding 3,164 independent mock DES surveys. To generate additional pseudo-independent DES Y3 shear mock maps, we rotated the four independent DES Y3 footprints by 45, 90, and 135 degrees in galactic longitude, thereby covering different (but overlapping) regions of the full-sky map. For each map we generated two distinct noise realizations, finally yielding a total of 25,312 pseudo-independent noisy DES Y3 mock maps. The maps from the first noise realization are used for the compression step, while those from the second noise realization are used for the cosmological inference step.

For this analysis, we augmented the Gower Street simulation suite with N-body simulations from the \texttt{DarkGridV1} suite \citep{Zuercher2021,Zuercher2021b}.  This integration was done because, unlike \paperA, we include $\Lambda$CDM results in this work, and the \texttt{DarkGridV1} simulations are exclusively $\Lambda$CDM simulations, in contrast to those from Gower Street. We later discover, however, that this addition has a negligible impact on our constraints (see Appendix~\ref{Appendix_dark_gower}), and the Gower Street simulations were sufficient also for the $\Lambda$CDM case. The \texttt{DarkGridV1} simulations suite explores 58 different $\Lambda$CDM cosmologies, varying $\Omega_{\textrm{m}}$ and $\sigma_8$. Each cosmology is represented by five independent full-sky simulations. For each full-sky simulation, we applied the same procedure as used for the Gower Street suite to generate multiple DES Y3 mock catalogues, resulting in an additional 2,320 mocks. These mocks are added to those from the Gower Street suite and are used both in the compression of the summary statistics and in the cosmological inference. 

In this analysis we only explicitly learn the likelihood surface for the cosmological parameters $\Omega_{\textrm{m}}$, $S_8$, $w$, and the intrinsic alignment amplitude $A_{\textrm{IA}}$; the dependence on other parameters is not explicitly learned, but is effectively marginalised over according to the distribution followed by the simulations (\S \ref{sect:LFI}). In contrast to the Gower Street suite, the \texttt{DarkGridV1} suite does not vary $h_{100}$, $n_s$, $\Omega_{\textrm{b}}$, or the neutrino mass. This implies that the distribution of these parameters when the two sets of simulations are combined is narrower than that arising from the Gower Street suite alone. We verified in Appendix \ref{Appendix_dark_gower} that this has a negligible impact on our posteriors.

%\NJ{A note on the effective prior change from using DarkGrid and, hopefully, a demonstration that it doesn't matter, would be needed here I think.}

Last, for testing purposes only, we also use a subset of the simulations from the \texttt{CosmoGridV1} suite \citep{cosmogrid1}. From this suite we chose a set of 100 full-sky simulations at the fiducial cosmology $\sigma_8 = 0.84$, $\Omega_{\textrm{m}}=0.26$, $w=-1$, $h=0.6736$, $\Omega_{\textrm{b}}=0.0493$, $n_{\textrm{s}}=0.9649$. Each simulation has been post-processed with a baryonification algorithm that mimics the impact of baryons at small scales. These simulations have been used only to produce the covariance matrices for the signal-to-noise estimates and to serve as mock data measurements for testing the full end-to-end pipeline. They have not been used for the compression of the summary statistics, nor during the cosmological inference.

The map-making procedure is detailed in \paperA{}. In brief, for each simulation we generate noisy mock shear maps following:
\begin{multline}
\label{eq:sc_pixel}
\gamma(p) = \frac{\sum_s \bar{n}(s) [1 + \GalaxyMatterBias{} \delta(p, s)] (1 + m) [\gamma(p, s)+\gamma_{\textrm{ IA}}(p, s)]}{\sum_s \bar{n}(s) [1 + \GalaxyMatterBias{} \delta(p, s)]}  + \\
\left(\frac{\sum_s \bar{n}(s)}{\sum_s \bar{n}(s) \left[1 + \GalaxyMatterBias{} \delta(p, s)\right]}\right)^{1/2} F(p) \, \frac{\sum_g w_g e_g}{\sum_g w_g}.
\end{multline}
Here $p$ is a simulation pixel; $s$ is a thin redshift shell; $\gamma(p, s)$ is the noiseless shear from the shear simulation; $\bar{n}(s)$ is the galaxy count across the whole footprint \citep{y3-sompz}; $m$ is the multiplicative shear bias that models shear measurement uncertainties \citep{y3-imagesims}; $\gamma_{\textrm{ IA}}(p, s)$ is the intrinsic alignment contribution to each pixel; $\delta(p, s)$ is the matter overdensity in the shear simulation; $\GalaxyMatterBias{}$ is the galaxy-matter bias of the weak lensing sample (fixed to unity); $w_g$ and $e_g$ are the DES Y3 galaxy weights and ellipticities for the galaxies $g$ in pixel $p$ (ellipticities have been randomly rotated to erase the cosmological signal of the catalogue); and $F(p)$ is a near-unity scale factor introduced to avoid double-counting source clustering effects (see \paperA{} for more details). The latter reads
\begin{equation}
    F(p) = A\sqrt{1-B \sigma_{e}^2(p)},
\end{equation}
where $A$ and $B$ are constants with values $A = [0.97,0.985,0.990,0.995]$ and $B = [0.1,0.05,0.035,0.035]$ (one for each tomographic bin), and  $\sigma_{e}^2(p)$ is the variance of the pixel noise. 
The intrinsic alignment term $\gamma_{\textrm{ IA}}(p, s)$ is:
\begin{equation}
    \label{eq:IA_}
    \gamma_{\textrm{ IA}}(p, s) = A_{\textrm{ IA}} \left(\frac{1+z}{1+z_0} \right)^{\eta_{\textrm{ IA}}} \frac{c_1 \rho_{m,0}}{D(z)} s(p,s),
\end{equation}
with $z_0= 0.62$, $c_1\rho_{m,0}=0.0134$ (\citealt{Bridle2007}), $D(z)$ the linear growth factor, and $s(p,s)$ the shear tidal field. We obtain $s(p,s)$ directly from the density field $\delta(p, s)$ by applying the (inverse) Kaiser-Squires algorithm.

%In our equation for intrinsic alignment, we've adapted the non-linear alignment (NLA) model but have extended it by incorporating source clustering effects from our simulations. This adds extra clustering terms to the intrinsic alignment model beyond what's found in the standard NLA. These additional terms resemble those in the tidal-torque alignment (TATT) model. However, while the TATT paper estimated these effects for catalogue-based Gaussian statistics through basic perturbation theory, our approach applies the simulation's clustering directly and extends it to all the summary statistics we're examining in this study.

This procedure is repeated for each of the four tomographic bins of the DES Y3 source catalogue. The noisy shear maps are then converted to noisy weak lensing mass maps using the same algorithm as for the data. The mock-making procedure has several free parameters: four multiplicative shear biases $m_i$ (one for each tomographic bin), four redshift distributions $n_i(z)$  (one for each tomographic bin), and parameters $A_{\textrm{IA}}$ and $\eta_{\textrm{ IA}}$ controlling the amplitude and the redshift evolution of intrinsic alignment. For each DES Y3 mock catalogue we draw at random one of these parameters from their prior (see Table \ref{parameter}); for the redshift distributions, we draw from one of the multiple realisations provided by \cite{y3-sompz}, which encompass the uncertainties in the redshift calibration of the DES Y3 $n(z)$.

\section{Summary statistics}
The summary statistics considered in this work are: 1) second and third moments; 2) wavelet phase harmonics; 3) the scattering transform. The summary statistics are applied to `smoothed' variants of the weak lensing maps, with the choice of smoothing varying according to the specific statistic employed: moments use top hat filters, while wavelet phase harmonics and the scattering transform use wavelet filters \citep{Cohen1995,Mallat1999,VDB1999}. We provide below a brief description of these statistics; for more details see \paperA{}.

\subsection{Second and Third Moments}\label{sect:2nd3rd}

Second moments of weak lensing mass maps are a Gaussian statistic, while third moments reflect the skewness of the field \citep{VanWaerbeke2013,Petri2015,Vicinanza2016,Chang2018,Vicinanza2018,Peel2018, G20,moments2021}.
%We adopt the same implementation of the second and third moments estimator implemented in \cite{moments2021}, who used second and third moments applied to DES y3 data to infer cosmology. 
Briefly, we first smooth the maps using top-hat filters, and we consider eight smoothing scales $\theta_0$ equally (logarithmically) spaced from $8.2$ to $221$ arcmin. The second and third moments estimators are:
\begin{equation}
\avg{\hat{\kappa}^2_{\theta_0}}(i, j) = \Avg_p \left( \kappa_{\theta_0,p}^i \, \kappa_{\theta_0,p}^j \right)
\end{equation}
\begin{equation}
\avg{\hat{\kappa}^3_{\theta_0}}(i, j, k) = \Avg_p \left( \kappa_{\theta_0,p}^i \, \kappa_{\theta_0,p}^j \, \kappa_{\theta_0, p}^k \right),
\end{equation}
where $\kappa_{\theta_0,p}^i$ is the smoothed lensing mass map of tomographic bin $i$ ($i,j,k$ refer to different tomographic bins), and the average is over all pixels $p$ on the full sky. In the case of third moments, we subtracted noise-signal third moments of the form  $\avg{\hat{\kappa}^{}_{\theta_0, \textrm{obs}} \, \hat{\kappa}^2_{\theta_0, \textrm{N}}}$; this approach was chosen because it improved the compression process and minimized the influence of source clustering on our summary statistics. 

\subsection{Wavelet Phase Harmonics}\label{sect:WPH} Wavelet phase harmonics (WPH) correspond to the second moments of smoothed weak lensing mass maps that have undergone a nonlinear transformation \citep{Cohen1995,Mallat1999,VDB1999,Allys2020}. A directional, multi-scale wavelet transform is used to smooth the maps. Let us consider a smoothed map $\kappa_{{n,\ell}}^i(\vec{\theta})$, where $n$ specifies the size of the filter (roughly equivalent to $2^{n+1}$ pixels), and $\ell$ the orientation. The non-linear operation on the smoothed map, which enables us to capture inter-scale interactions and access the non-Gaussian characteristics of the field through second moments, is defined as:
\begin{equation}
    \PH(re^{\imagunit{} \theta},q) \equiv r e^{\imagunit{} q \theta},
\end{equation}
where $r$ is the modulus of the field and $\theta$ its phase. We consider only $q=0$ or $q=1$ in this work, corresponding to taking the modulus or leaving the field unaltered, respectively. The WPH statistics used in this work are:
\begin{equation}
    S00(i, j, n) =  \Avg_p \Avg_{\ell} \left( |\kappa_{{n,\ell}}^i| \, |\kappa_{{n,\ell}}^j|\right)
\end{equation}
\begin{equation}
    S11(i, j, n) \equiv \WPHG(i, j, n) = \Avg_p \Avg_{\ell} \left( \kappa_{{n,\ell}}^i \, \kappa_{{n,\ell}}^j \right)
\end{equation}
\begin{equation}
    S01(i, j, n) = \Avg_p \Avg_{\ell} \left( |\kappa_{{n,\ell}}^i| \,\, \kappa_{{n,\ell}}^j \right)
\end{equation}
\begin{equation}
    C01{\delta \ell 0}(i, j, n_1, n_2) =  \Avg_p \Avg_{\ell} \left( |\kappa_{{n_1,\ell}}^i| \,\, \kappa_{{n_2,\ell}}^j \right) \textrm{ for } n_1<n_2
\end{equation}
\begin{equation}
    C01{\delta \ell 1}(i, j, n_1, n_2) = \Avg_p \Avg_{\ell} \left( |\kappa_{{n_1,\ell+1}}^i| \,\, \kappa_{{n_2,\ell}}^j \right) \textrm{ for } n_1<n_2.
\end{equation}
We average over all pixels $p$ and we also average over the three values $0,1,2$ of the rotation index $\ell$ (corresponding to three possible orientations of the directional wavelet). The number $n$ varies from $0$ to $5$. The above summary statistics probe both Gaussian and non-Gaussian features of the field (with the exception of $\WPHG$, which is equivalent to the power spectrum of $\kappa$). To reduce the impact of source clustering, for WPH $S01$, $C01{\delta \ell 0}$, and $C01{\delta \ell 1}$ we subtract a term involving one noise-only map and the observed noisy convergence map.

\subsection{Scattering Transform}\label{sect:ST}

The scattering transform \cite{Mallat2012,Bruna2013,Cheng2020,Valogiannis2022a,Valogiannis2022} conceptually resembles the wavelet phase harmonics (WPH) previously introduced. It involves smoothing the field using a directional, multi-scale wavelet transform (same as implemented in WPH), but then followed by a modulus operation. This transform-then-modulus operation is iteratively applied $m$ times, after which we take an overall field average; the result is the scattering transform coefficient $\text{ST}_m$. This study considers scattering coefficients of orders $m=1,2$. Given a directional multi-scale wavelet $\psi_{n,\ell}$ and the convergence map $\kappa^i$ of tomographic bin $i$, we obtain:
\begin{equation}
    \ST_{1}(i, n) = \Avg_p \Avg_{\ell} \left( | \kappa^i  \ast \psi_{n,\ell} | \right)
\end{equation}
\begin{multline}
    \ST_{2}(i, n_1, n_2, \ell') =  \Avg_p \Avg_{\ell} \left( | \, | \kappa^i  \ast \psi_{n_1,\ell} | \ast \, \psi_{n_2,\ell'-\ell} | \right)\\ \textrm{ for } n_1 \le n_2,
\end{multline}
where the average runs over all the pixels $p$ and over all values of the rotation index $\ell$ (similar to the WPH case).

\begin{figure}
\includegraphics[width=0.48\textwidth]{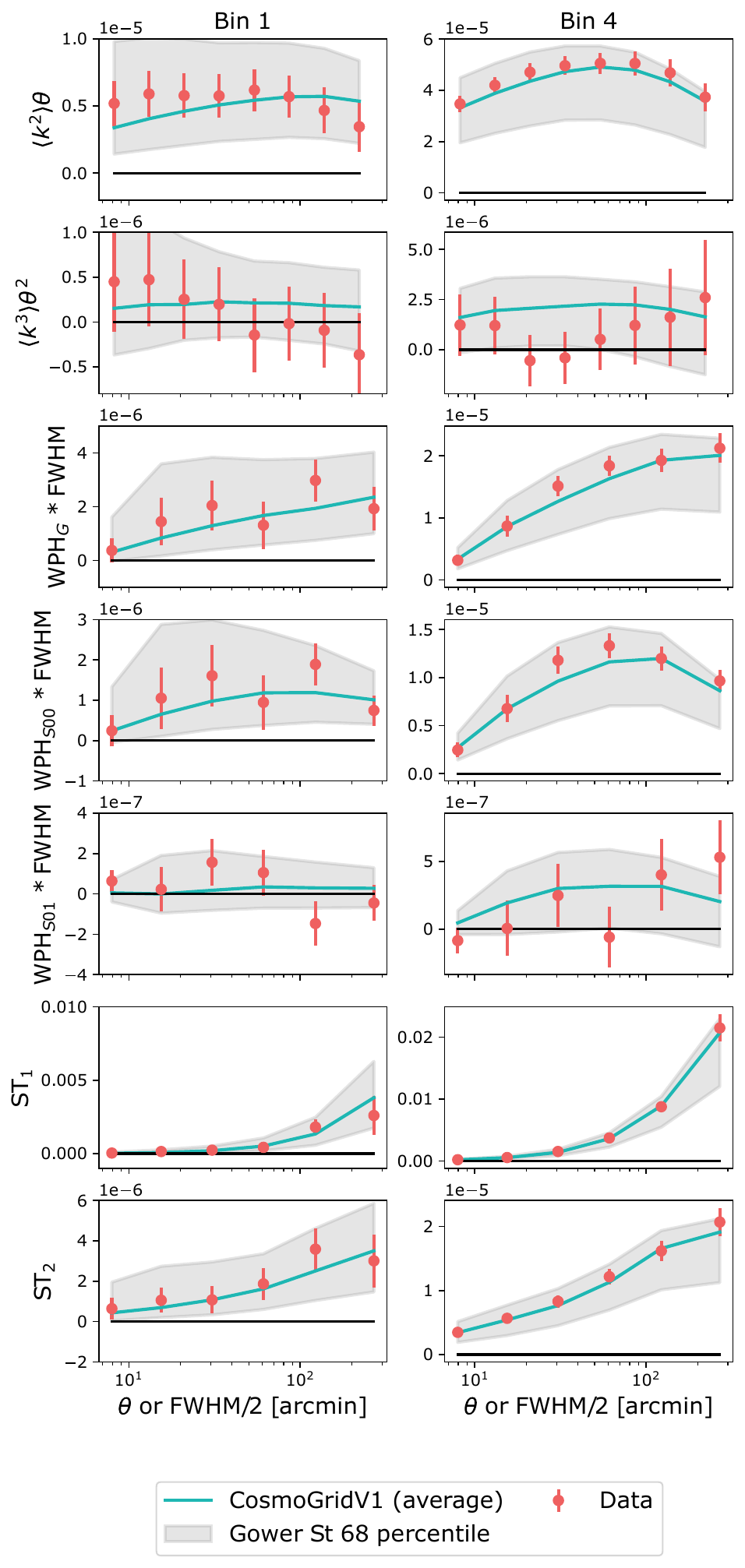}
\caption{Some of the Gaussian and non-Gaussian statistics considered in this work, for the first (left column) and fourth (right column) tomographic bins. Red points are the statistic as measured in data, using the noisy convergence map; error bars have been estimated using 400 measurements at a fixed cosmology from the \texttt{CosmoGridV1} simulations \citep{cosmogrid1}. Blue lines show the measurements in the \texttt{CosmoGridV1} simulations, averaged over the multiple simulations available at the fiducial cosmology. The blue lines only serve to guide the eye, as they are not a best fit to our data. However, the cosmological parameters of the \texttt{CosmoGridV1} simulations are not too dissimilar from our best-fit parameters, so the blue lines provide a reasonable fit. For statistics involving filters with different sizes $j_1,j_2$, we considered $j_1=j_2$. The grey region shows for comparison the 68 \% contour of the noisy measurements from the Gower Street simulations with $|A_{\textrm{ IA}}| < 1$ and $|\eta_{\textrm{ IA}}| < 1$ (which correspond to $\sim$ 500 mock measurements).}
\label{fig:DV}
\end{figure}

\subsection{Measurement in data and signal-to-noise}

\renewcommand{\arraystretch}{1.1} % Adjusts the row height 
\begin{table*}
%\tiny
\caption {Salient properties of the summary statistics. The second column denotes whether it carries Gaussian (G) or non-Gaussian (NG) information. The third column refers to the order of the field $\kappa$. The fourth column is the number of components of the data vector across scales and tomographic bins. The further columns show the signal-to-noise ratio (SN) of the measurements in data.}
\centering
%\begin{adjustbox}{width=0.7\textwidth}
\begin{tabular}{l c c c c c c c}
 \toprule
& \textbf{G/NG} & \textbf{Order} & \textbf{Data vector length} & \textbf{SN(Bin 1)} & \textbf{SN(Bin 2)} & \textbf{SN(Bin 3)} & \textbf{SN(Bin4)} \\
 \midrule
2nd moments & G & 2  &  160     & 3.9 & 7.8 & 16.6 & 13.7\\
3rd moments & NG & 3 &  512    & 0.7 & 1.3 & 2.5 & 1.9\\
WPH S11 ($\WPHG$) & G & 2 & 120   & 3.4 & 7.1 & 15.9 & 13.1\\
WPH S00 & NG & 2 & 96        & 3.2 & 6.7 & 15.2 & 12.3\\
WPH S01 & NG & 2 & 480        & 0.5 & 2.5 & 2.1 & 3.2\\
ST1  & NG & 1 & 60         & 3.4 & 7.6 & 16.3 & 14.4\\
ST2  & NG & 1 & 630         & 3.3 & 7.3 & 15.9 & 13.7\\
\bottomrule
\end{tabular}
%\end{adjustbox}
\label{scalesetc}
\end{table*}

We present various statistics in Fig. \ref{fig:DV}, both as measured in the data and as measured in a set of simulations that we did not use for our cosmological inference pipeline (CosmoGridV1 simulations \citep{cosmogrid1}). The simulated data vector is used solely to guide the eye, as the simulations are not expected to provide the best fit to the measurements in the data. The figure also shows the $68\%$ contour spanned by the noisy measurements from the Gower Street simulations; these encompass the measurements in data very effectively. When calculating this region, we limited the analysis to an Intrinsic Alignment (IA) interval smaller than the Gower Street prior ($|A_{\textrm{ IA}}| < 1$ and $|\eta_{\textrm{ IA}}| < 1$) to simplify visualisation. Without this restriction, the grey region in the first bin would be significantly larger, as it can be highly impacted by large IA values. We also report the measured signal-to-noise ratio in Table \ref{scalesetc}; these ratios are very similar to the ones reported in \paperA{} for simulated measurements.

\subsection{Data Compression}

%\NJ{Needs a note on which simulations were used to train, which for validation, and which for DELFI. }

Data compression increases the efficiency of density estimation by reducing the data vector's dimensionality. In this work we use neural compression and we match the dimension of the compressed summary statistics to that of the parameters of interest $\theta$.
Specifically, for a summary statistic $\mathbf{d}$, we compress it as $\mathbf{t} = F_{\phi}(\mathbf{d})$, modelling the compression function $F_{\phi}$ with a neural network. We optimize $\phi$ by minimizing a mean squared error (MSE) loss using the first half of our pseudo-independent mocks.  The architecture used for the network and the number of parameters are summarised in Table 3 in \paperA{}.

Given that we are principally focused on constraining the parameters $\Omega_{\textrm{m}}$, $S_8$, $w$, and $A_{\textrm{IA}}$, we target these parameters exclusively for individual compression and refrain from compressing the data vectors for any other parameters. We compress each summary statistic separately and then amalgamate their compressed forms during the likelihood inference stage. Specifically, we compress the second moments, third moments, $\WPHG$, WPH $S00$, $\ST_1$, and $\ST_2$ on an individual basis, whereas for optimisation purposes we compress WPH $S01$ and WPH $C01$ jointly.

%We optimize $\phi$ by minimizing a mean squared error (MSE) loss using the first half of our pseudo-independent mocks. The architecture used for the network and the number of parameters are summarised in Table 3 in \paperA{}.  %As we are mostly interested in  the constraints on $\Omega_{\textrm{m}}$, $S_8$, $w$ and $A_{\textrm{IA}}$, we use these four parameters for the compression. 

\section{Simulation-based inference and posterior estimation}\label{sect:LFI}

This work relies on simulation-based inference (also known as likelihood-free inference) to ensure a reliable inference of the cosmological parameters. More details are given in \paperA{}; in brief, in simulation-based inference the likelihood is not assumed to have a closed form, but rather is estimated from mock noisy realisations of the summary statistics. We use an ensemble of neural density estimators (NDEs) -- Gaussian Mixture Density Networks (MDNs; \cite{Bishop1994}) and Masked Autoregressive Flows (MAF; \cite{Papamakarios2017}) --  to estimate the conditional distribution $p(d|\theta)$. NDEs approximate this distribution with estimates $q(d|\theta; \phi')$ where the network parameters $\phi'$ are optimized by minimizing a loss function.

We use the package pyDELFI \citep{Alsing2018} both for density estimation and for neural network training. For the training phase, we input the compressed statistics from half of our simulations, specifically those not used in the compression step. Density estimation is performed using the parameter set $\theta = \left[ \Omega_{\textrm{m}}, S_8, w, A_{\textrm{ IA}}\right]$ and the corresponding compressed data vectors. This approach implies that the remaining parameters, which vary in the mock productions, are effectively marginalized over \citep{momentnets}. The process of marginalization respects the prior distributions that were applied to sample these parameters during the mock generations, as detailed in Table~\ref{parameter}.

The final density estimation combines the different ensemble estimates (MDNs and MAFs) weighted by training-derived losses. With the estimated target distribution $p(d|\theta)$, we compute the likelihood at observed data $d=d_{\textrm{obs}}$.  The final posteriors are derived via Markov chain Monte Carlo (MCMC) sampling of the likelihood, while considering the priors mentioned in Table \ref{parameter} for the parameters $\theta = \left[ \Omega_{\textrm{m}}, S_8, w, A_{\textrm{ IA}}\right]$. For the $\Lambda$CDM analysis, we simply run the MCMC sampling fixing $w=-1$. This MCMC sampling is carried out using the publicly available software package \texttt{EMCEE} \citep{Foreman-Mackey2013}, an ensemble sampler with affine-invariant properties designed for MCMC. %In \paperA{} we tested that the posteriors

%In the likelihood -free inference (also know as simulation-based inference), the likelihood $p(d|\theta)$ is not assumed to have a closed form, but it is rather reconstructed from simulated mock data as part of the inference pipeline. See \paperA{} for a summary of the procedure we followed to infer the parameters posterior; a more detailed description is provided in \cite{jeffrey_lfi}.

\section{Pre-unblinding tests}\label{sect:unblinding}

Blinding procedures are commonly used in weak lensing analyses to prevent observer biases. Researchers intentionally conceal from themselves specific details of the data or analysis pipeline until late in the analysis; this promotes objectivity and reduces the influence of expectations on the results. The main DES Y3 weak lensing analysis \citep{y3-cosmicshear1,y3-cosmicshear2} uses a two-stage blinding scheme. First, the weak lensing sample was blinded by means of a multiplicative factor \citep{y3-shapecatalog}; second, the summary statistics under study were manipulated according to a transformation introduced by \cite{y3-blinding}. This transformation was specifically designed to induce a shift in the posterior distributions, while ensuring that the measured data vectors could still be subjected to systematic tests and analysis. At the time of writing, the shape catalogue was already published and unblinded; therefore the first level of blinding was not applied in this paper. Furthermore, our pipeline does not accommodate the second blinding stage, as that stage relies on a parameter-based model of noise-free summary statistics, which we lack. Given that the main DES Y3 cosmological findings are already public, we opted not to blind our catalogue or data vectors. Instead, we refrained from examining the data posteriors until after our pipeline validation and null tests on the data vectors were successfully completed. In addition to the tests conducted at the catalogue and map level, as described in \cite{y3-shapecatalog} and \cite*{y3-massmapping}, we performed the following tests, detailed in \paperA{}:
\begin{itemize}
    
    \item \textbf{Validation of the Gower Street simulations}: We tested that the power spectra of the noiseless, full-sky convergence of the Gower Street simulations were in an approximately one percent agreement with theory predictions; moreover, we tested that the noise properties of our mocks (i.e. second, third, and fourth moments of the noise, as well as its cumulant distribution function) were able to describe the noise properties of our data.
    
    \item \textbf{Validation of the posteriors}: We checked, using an empirical coverage test, that the size of the posteriors estimated by our NDEs was not misestimated.
    
    \item \textbf{End-to-end validation of the pipeline on simulations}: We tested, using an independent set of simulations, that the pipeline was able to recover true cosmology for the summary statistics considered in this work.

   % \item \textbf{End-to-end comparison of the pipeline against a `standard', theory-based pipeline}: We compared the cosmological constraints obtained from a simulated power spectrum using the LFI pipeline against a more standard approach were we relied on a theoretical model for the power spectrum and we assume the Likelihood to be Gaussian.
    
    \item \textbf{Tests on baryonic contamination}: We conducted tests using a subset of simulations that have been post-processed to incorporate baryonic feedback via the baryonic correction model \citep{Schneider2015,Arico2020}. Our objective was to verify that our statistics and the scales utilized in this study were not significantly impacted by potential baryonic contamination, which the simulations in our pipeline do not model. Specifically, we confirmed that the peak of the marginalized two-dimensional posterior distribution of $\Omega_{\textrm{m}}$ and $S_8$, when analyzing baryon-contaminated data, falls within $0.3\sigma$ of the peak obtained from clean data.
    
    \item \textbf{Tests on additive biases due to PSF modelling errors}: We checked that additive biases due to PSF modelling errors were negligible at the data vector level, i.e. that if neglected they would not bias our cosmological analysis. This test is similar to that performed in the DES Y3 cosmic shear analysis \citep{y3-cosmicshear1}.
    
    \item \textbf{Tests on possible modelling errors of the source clustering effect}: We checked that potential errors in our model for the source clustering effects were negligible, i.e. that they would not bias our cosmological analysis.

\end{itemize}

These tests were performed in \paperA{} for a $w$CDM analysis. Appendix \ref{ref:validation_LCDM} repeats some of these for a $\Lambda$CDM analysis, and also adds two sets of tests:

\begin{itemize}
    
    \item \textbf{Test on the impact of redshift uncertainties and shear biases}: We show that we can recover the true cosmology of the simulations even if the parameters describing redshift uncertainties and shear bias calibrations were $2\sigma$ off their mean values.
    
    \item \textbf{Sensitivity to the details of the N-body simulations}: We show that our {posteriors are not sensitive to the details of the simulations used} (box size, number of particles, redshift resolution).

\end{itemize}

Finally, before examining the posteriors we performed three additional tests:

\begin{itemize}
\item \textbf{B-mode `null-test'}: Weak lensing does not produce B-modes at first order, making such modes useful for identifying systematics. For masked data, however, the map-making procedure can cause small B-modes due to E-modes leakage at large scales. We perform in Appendix \ref{ref:B_modes} a comparison between data and simulated B-modes; this suggests no anomaly in our data measurements.

%\NJ{Again a note that this is for the compressed data vectors, t}

\item \textbf{Goodness-of-fit}: We assess the goodness-of-fit $p$-value for the compressed data vectors, ensuring they do exceed 1 percent ($p$-value$>0.01$, see Section~\ref{sect:results}, Tables \ref{final_results_wcdm},  \ref{final_results_lcdm}). We used the following procedure to measure $p$-values. Let $\mathcal{L}(\mathbf{x})$ denote the likelihood of a data vector $\mathbf{x}$. For each summary statistic (or combination of statistics), we trained the Neural Density Estimators on all (compressed) simulated data vectors except one, denoted $\mathbf{x}_i$, selected at random. We then estimated the minimum log likelihood\footnote{The routine \texttt{scipy.optimize.differentialevolution} was used.} from the learned likelihood surface for the simulated data vector excluded from the training; we denote this minimum log likelihood by
\begin{equation}
\text{minLL}_i \equiv \min \log \mathcal{L}(\mathbf{x}_i).
\end{equation}
This process was repeated for a thousand data vectors. Next, we estimated the minimum log likelihood for our data:
\begin{equation}
\text{minLL}_{\text{data}} = \min \log \mathcal{L}(\mathbf{d}).
\end{equation}
Subsequently, we computed the probability-to-exceed $p$ by counting the fraction of simulated data vectors that had a minimum log likelihood greater than that of the observed data:
\begin{equation}
p = \frac{1}{N} \sum_{i=1}^{N} \vmathbb{1}(\text{minLL}_i > \text{minLL}_{\text{data}}),
\end{equation}
where $\vmathbb{1}$ is the indicator function, which is 1 if its argument is true and 0 otherwise. We note that this goodness-of-fit test is valid only for compressed data vectors; a satisfactory $p$-value for compressed statistics does not necessarily imply a similar outcome for uncompressed statistics. Unfortunately, we are unable to test the uncompressed statistics, as learning the likelihood of the uncompressed measurements is not feasible with the limited simulations available to us. However, we note that the uncompressed data are effectively encompassed by the uncompressed mock measurements from the Gower Street simulations, as depicted in Fig. \ref{fig:DV}. This provides evidence that the uncompressed data vector does not exhibit features that are unaccounted for in our mock simulations.

\item \textbf{Stability of the posterior}: We visually checked that the four NDEs were delivering consistent posteriors when considered individually (Appendix \ref{sect:individuals}). This was done by blinding the cosmological parameter axes.

\item \textbf{Internal consistency of summary statistics}: We assess the internal consistency of the various summary statistics employed. To do this, we use the Posterior Predictive Distribution (PPD) methodology, developed by \cite{Doux2021} and implemented in the main DES Y3 analysis. This methodology is designed to calculate a calibrated probability-to-exceed $p$-value. For consistency tests between two types of summary statistics, we generate realisations of one type of summary statistic (e.g. second moments) under the assumption that the other type (e.g. third moments) is the one measured in data. These realisations are obtained using parameters drawn from the posterior combining both types of statistics. In our case, we easily obtained these realisations from the joint likelihood,  which we learned using the noisy (compressed) measurements from the Gower Street simulations. These realisations are then compared to actual observations using a distance metric (in our case, the log likelihood) in the compressed data space; this metric is then used to determine the $p$-value. More details on the exact implementation are given in \cite{Doux2021}.  We report in Table \ref{table:PPD} the PPD $p$-values obtained comparing second moments against third moments, ST, and WPH (and vice-versa); all the values are well above the $p=0.01$ threshold. Note that the PPD calculation is not symmetric in its two arguments.

\end{itemize}

Only once all these tests were passed did we look at the unblinded posteriors of our analysis.

\section{Results}\label{sect:results}

\renewcommand{\arraystretch}{1.5} % Adjusts the row height of the table
\begin{table*}
%\tiny
\caption {Constraints on various parameters for different summary statistics and their combinations, for the $\Lambda$CDM model. For each parameter we report the mean and 68 percent credible interval. The last two columns report the FoM and the $p$-values.}
\centering
%\begin{adjustbox}{width=0.7\textwidth}
\begin{tabular}{l c c c c c c}
 \toprule
\textbf{$\Lambda$CDM Summary Statistic(s)} & $S_8$ &  $\sigma_8$ & $\Omega_{\textbf{m}}$  & $A_{\textrm{IA}}$ & ${\textrm{FoM}(S_8,\Omega_{\textrm{m}})}$ & $p$-value\\
 \midrule

%2nd moments &0.796 $\pm$ 0.0204 &0.849 $\pm$  0.057  &0.267 $\pm$  0.033   &0.47 $\pm$  0.45  &  1303&  0.39\\
%2nd+3rd moments &0.801 $\pm$ 0.0208&0.847  $\pm$ 0.047&0.270 $\pm$ 0.029&0.74 $\pm$ 0.51& 1398 & 0.79\\
%2nd moments+ST &0.791 $\pm$ 0.0186&0.846  $\pm$ 0.046&0.264 $\pm$ 0.027&0.50 $\pm$ 0.34& 1770 & 0.41\\
%2nd moments+WPH &0.801 $\pm$ 0.0198&0.866  $\pm$ 0.045&0.259 $\pm$ 0.027&0.50 $\pm$ 0.44& 1817 & 0.75\\
%2nd moments+ST+WPH &0.791 $\pm$ 0.0178&0.855  $\pm$ 0.044&0.259 $\pm$ 0.027&0.47 $\pm$ 0.36& 2069 & 0.60\\
%2nd+3rd moments+ST+WPH &0.794 $\pm$ 0.0172&0.857  $\pm$ 0.042&0.259 $\pm$ 0.025&0.45 $\pm$ 0.32& 2234 & 0.71\\

2nd moments &0.796$^{+0.021}_{-0.020}$&0.849$^{+0.055}_{-0.056}$&0.267$^{+0.033}_{-0.033}$&0.45$^{+0.48}_{-0.42}$& 1303 & 0.39\\
2nd+3rd moments &0.801$^{+0.021}_{-0.021}$&0.847$^{+0.044}_{-0.044}$&0.270$^{+0.028}_{-0.029}$&0.51$^{+0.53}_{-0.50}$& 1398 & 0.79\\
2nd moments+ST &0.791$^{+0.019}_{-0.018}$&0.846$^{+0.044}_{-0.044}$&0.264$^{+0.027}_{-0.028}$&0.34$^{+0.35}_{-0.33}$& 1770 & 0.41\\
2nd moments+WPH &0.801$^{+0.020}_{-0.020}$&0.866$^{+0.044}_{-0.045}$&0.259$^{+0.027}_{-0.026}$&0.44$^{+0.45}_{-0.43}$& 1817 & 0.75\\
2nd moments+ST+WPH &0.791$^{+0.018}_{-0.018}$&0.855$^{+0.043}_{-0.044}$&0.259$^{+0.027}_{-0.027}$&0.36$^{+0.37}_{-0.36}$& 2069 & 0.60\\
2nd+3rd moments+ST+WPH &0.794$^{+0.017}_{-0.017}$&0.857$^{+0.042}_{-0.042}$&0.259$^{+0.025}_{-0.025}$&0.32$^{+0.33}_{-0.31}$& 2234 & 0.71\\

\midrule
%DES Y3 cosmic shear  &0.788 $\pm$ 0.017&0.810  $\pm$ 0.066&0.289 $\pm$ 0.041&0.33 $\pm$ 0.28& 1402 & \\

DES Y3 cosmic shear  &0.788$^{+0.016}_{-0.019}$&0.810$^{+0.061}_{-0.071}$&0.289$^{+0.044}_{-0.039}$&0.33$^{+0.28}_{-0.27}$& 1402 & \\

%DES Y3 2nd+3rd moments (G22)  &0.802 $\pm$ 0.019&0.820  $\pm$ 0.033&0.304 $\pm$ 0.031&0.08 $\pm$ 0.49& 1747 & \\

DES Y3 2nd+3rd moments (G22)   &0.787$^{+0.021}_{-0.016}$&0.819$^{+0.038}_{-0.028}$&0.279$^{+0.033}_{-0.029}$&0.40$^{+0.61}_{-0.38}$& 1747 & \\

%DES Y3 Cl+Peaks (Z22)  &0.797 $\pm$ 0.014&0.849  $\pm$ 0.108&0.276 $\pm$ 0.071&-0.03 $\pm$ 0.19& 1031 & \\

DES Y3 Cl+Peaks (Z22)  &0.797$^{+0.014}_{-0.014}$&0.849$^{+0.100}_{-0.117}$&0.276$^{+0.077}_{-0.064}$&-0.04$^{+0.19}_{-0.19}$& 1031 & \\
%DES Y3 Cl+Peaks (Z22)  &0.797 $\pm$ 0.013&0.847  $\pm$ 0.081&0.278 $\pm$ 0.051&-0.03 $\pm$ 0.18& 1313 &
 \bottomrule
\end{tabular}
%\end{adjustbox}
\label{final_results_lcdm}
\end{table*}

\renewcommand{\arraystretch}{1.6} % Adjusts the row height of the table
\begin{table*}
%\tiny
\caption {Constraints on various parameters for different summary statistics and their combinations, for the $w$CDM model. For each parameter we report the mean and the 68 percent credible interval (except that for $w$ we report the mean and 68 percent upper limit). The last two columns report the FoM and the $p$-values.}
\centering
%\begin{adjustbox}{width=0.7\textwidth}
\begin{tabular}{l c c c c c c c}
 \toprule
\textbf{$w$CDM Summary Statistic(s)} & $S_8$ &  $\sigma_8$ & $\Omega_{\textbf{m}}$  & $w$ & $A_{\textrm{IA}}$ & ${\textrm{FoM}(S_8,\Omega_{\textrm{m}})}$ & $p$-value\\
 \midrule
%2nd moments &0.826 $\pm$ 0.028 & 0.870 $\pm$  0.059  & 0.274 $\pm$  0.037   & < -0.62   &0.31 $\pm$  0.33  &  899&  0.66\\
%2nd+3rd moments &0.825 $\pm$ 0.026 &0.858 $\pm$  0.049  &0.280 $\pm$  0.033   & < -0.69   &0.58 $\pm$  0.39  &  1121&  0.73\\
%2nd moments+ST &0.814 $\pm$ 0.024 &0.847 $\pm$  0.047  &0.279 $\pm$  0.032   & < -0.68   &0.38 $\pm$  0.29  &  1274&  0.30\\
%2nd moments+WPH &0.830 $\pm$ 0.024 &0.862 $\pm$  0.044  &0.281 $\pm$  0.032   & < -0.65   &0.35 $\pm$  0.30  &  1435&  0.90\\
%2nd moments+ST+WPH &0.819 $\pm$ 0.021 &0.855 $\pm$  0.045  &0.277 $\pm$  0.032   & < -0.70   &0.28 $\pm$  0.25  &  1605&  0.58\\
%2nd+3rd moments+ST+WPH &0.817 $\pm$ 0.021 &0.861 $\pm$  0.042  &0.273 $\pm$  0.029   & < -0.72   &0.37 $\pm$  0.29  &  1725&  0.67\\

2nd moments &0.826$^{+0.028}_{-0.028}$&0.870$^{+0.056}_{-0.057}$&0.274$^{+0.037}_{-0.037}$& < -0.62 &0.31$^{+0.37}_{-0.29}$& 899 & 0.66\\
2nd+3rd moments &0.825$^{+0.026}_{-0.027}$&0.858$^{+0.047}_{-0.047}$&0.280$^{+0.033}_{-0.033}$& < -0.69 &0.58$^{+0.39}_{-0.39}$& 1121 & 0.73\\
2nd moments+ST &0.814$^{+0.024}_{-0.024}$&0.847$^{+0.046}_{-0.046}$&0.279$^{+0.032}_{-0.033}$& < -0.68 &0.38$^{+0.30}_{-0.29}$& 1274 & 0.30\\
2nd moments+WPH &0.830$^{+0.024}_{-0.024}$&0.862$^{+0.043}_{-0.043}$&0.281$^{+0.032}_{-0.032}$& < -0.65 &0.35$^{+0.31}_{-0.30}$& 1435 & 0.90\\
2nd moments+ST+WPH &0.819$^{+0.021}_{-0.021}$&0.855$^{+0.045}_{-0.045}$&0.277$^{+0.032}_{-0.032}$& < -0.70 &0.28$^{+0.25}_{-0.25}$& 1605 & 0.58\\
2nd+3rd moments+ST+WPH &0.817$^{+0.021}_{-0.021}$&0.861$^{+0.041}_{-0.041}$&0.273$^{+0.029}_{-0.029}$& < -0.72 &0.37$^{+0.30}_{-0.29}$& 1725 & 0.67\\

\midrule
%DES Y3 cosmic shear  &0.813 $\pm$ 0.027&0.816  $\pm$ 0.080&0.303 $\pm$ 0.045& < -0.77 &0.34 $\pm$ 0.26& 901 & \\

DES Y3 cosmic shear  &0.813$^{+0.032}_{-0.022}$&0.816$^{+0.082}_{-0.077}$&0.303$^{+0.040}_{-0.051}$& < -0.77 &0.34$^{+0.28}_{-0.25}$& 901 & \\
\bottomrule
\end{tabular}
%\end{adjustbox}
\label{final_results_wcdm}
\end{table*}
\renewcommand{\arraystretch}{1.1} % Adjusts the row height of the table

\begin{table}
    \centering
    \begin{tabular}{l c}
        \toprule
        \textbf{$\Lambda$CDM Data Splits} &  \textbf{$p$-values}  \\
        \midrule
        2nd moments \textit{vs.} 3rd moments& 0.50 \\
        3rd moments \textit{vs.} 2nd moments& 0.20 \\
        2nd moments \textit{vs.} ST&  0.58 \\
        ST \textit{vs.} 2nd moments&  0.28 \\
        2nd moments \textit{vs.} WPH& 0.16 \\
        WPH \textit{vs.} 2nd moments& 0.11 \\
        \bottomrule
        \\
        \toprule
        \textbf{$w$CDM Data Splits} &  \textbf{$p$-values}  \\
        \midrule
        2nd moments \textit{vs.} 3rd moments& 0.47 \\
        3rd moments \textit{vs.} 2nd moments& 0.36 \\
        2nd moments \textit{vs.} ST&  0.45 \\
        ST \textit{vs.} 2nd moments&  0.33 \\
        2nd moments \textit{vs.} WPH& 0.14 \\
        WPH \textit{vs.} 2nd moments& 0.13 \\
        \bottomrule        
    \end{tabular}
    \caption{Summary of internal consistency test $p$-values. All internal consistency tests pass the pre-defined (arbitrary) threshold of 0.01.}
\label{table:PPD}
\end{table}

\begin{figure*}
\begin{center}
\includegraphics[width=0.46\textwidth]{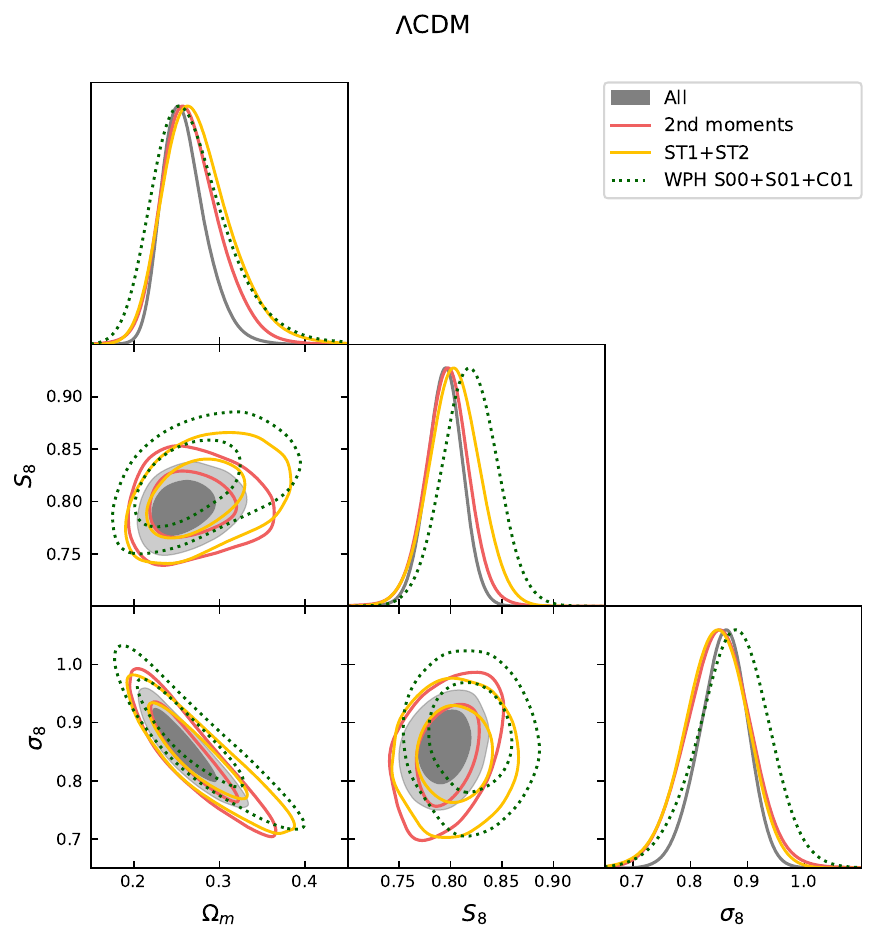}
\includegraphics[width=0.45\textwidth]{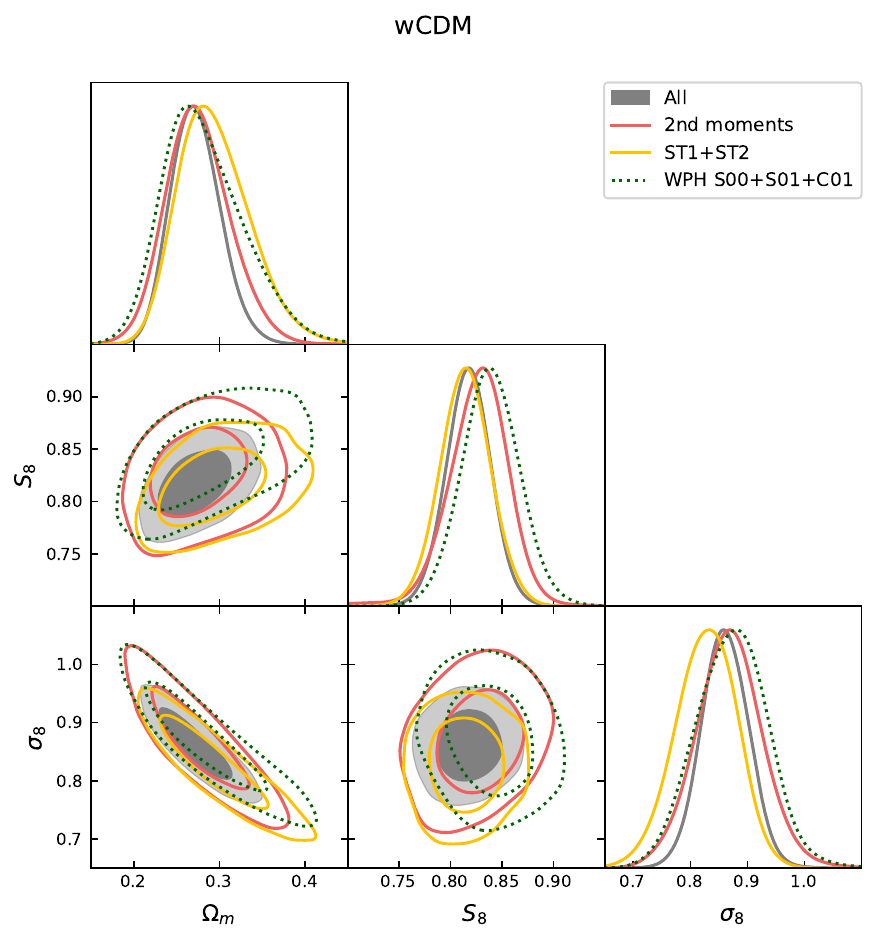}
\end{center}
\caption{Posterior distributions of the cosmological parameters $\Omega_{\textrm{m}}$, $S_8$, and $\sigma_8$, for different summary statistics and their combinations (``All''), as measured in our data ($\Lambda$CDM  on the left, $w$CDM  on the right). The two-dimensional marginalised contours in these figures show the 68 percent and 95 percent credible regions.}
\label{fig:all}
\end{figure*}

\begin{figure}
\begin{center}
\includegraphics[width=0.45\textwidth]{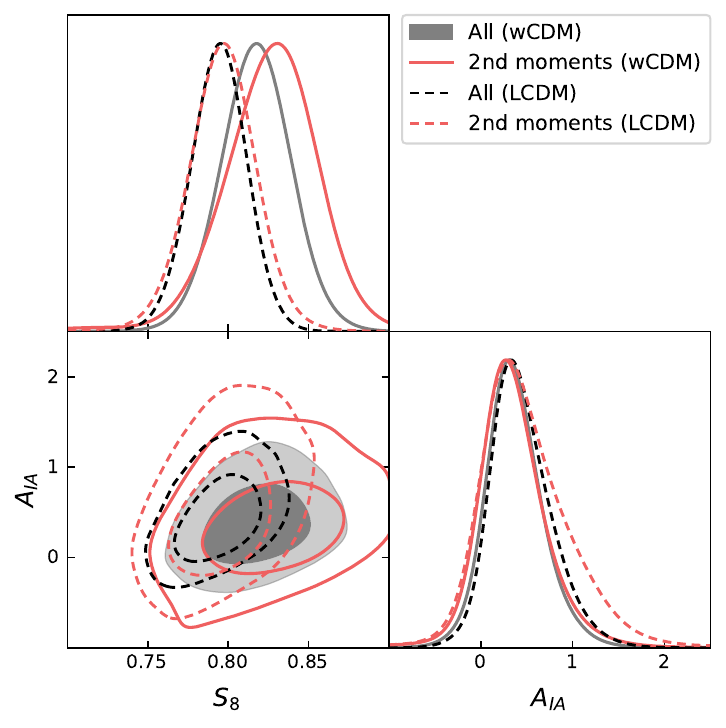}
\end{center}
\caption{Posterior distributions of the cosmological parameters $S_8$ and $A_{\textrm{IA}}$, for second moments and for the combination of all the summary statistics (``All''), to better highlight the contribution of non-Gaussian statistics to the overall constraints. The two-dimensional marginalised contours in these figures show the 68 percent and 95 percent credible regions.}
\label{fig:all_IA}
\end{figure}

\begin{figure}
\begin{center}
\includegraphics[width=0.45\textwidth]{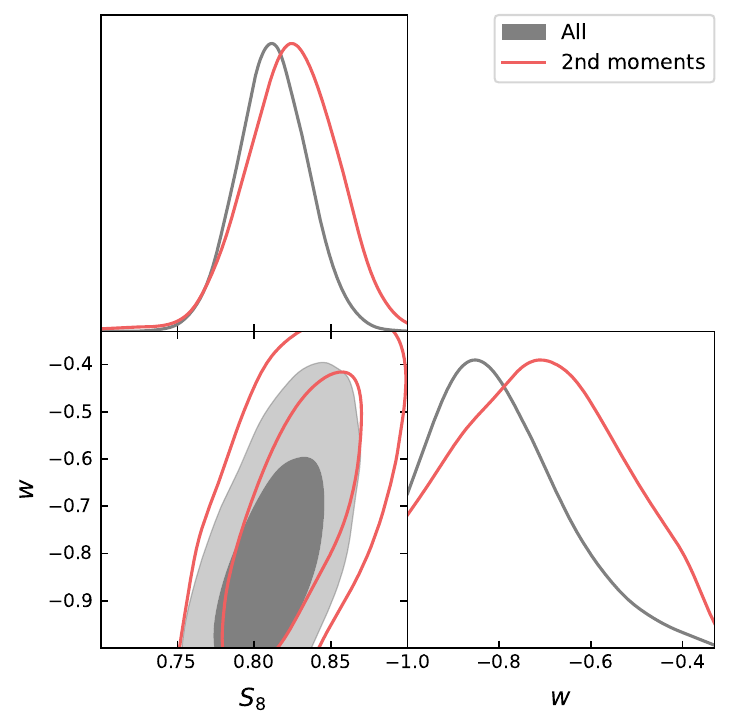}
\end{center}
\caption{Posterior distributions of the cosmological parameters $S_8$ and $w$, for second moments and for the combination of all the summary statistics (``All''), to better highlight the contribution of non-Gaussian statistics to the overall constraints.%\LW{Why are not other combinations of statistics shown?} 
The two-dimensional marginalised contours in these figures show the 68 percent and 95 percent credible regions.}
\label{fig:all_w}
\end{figure}

\begin{figure*}
\begin{center}
\includegraphics[width=0.45\textwidth]{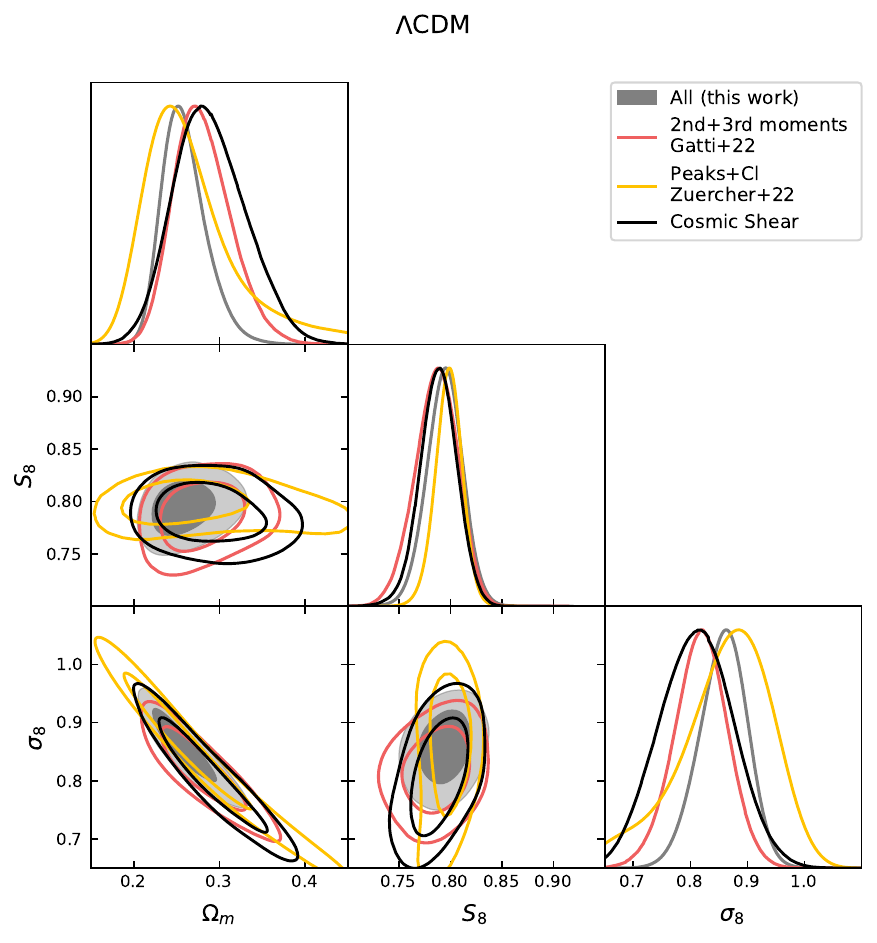}
\includegraphics[width=0.45\textwidth]{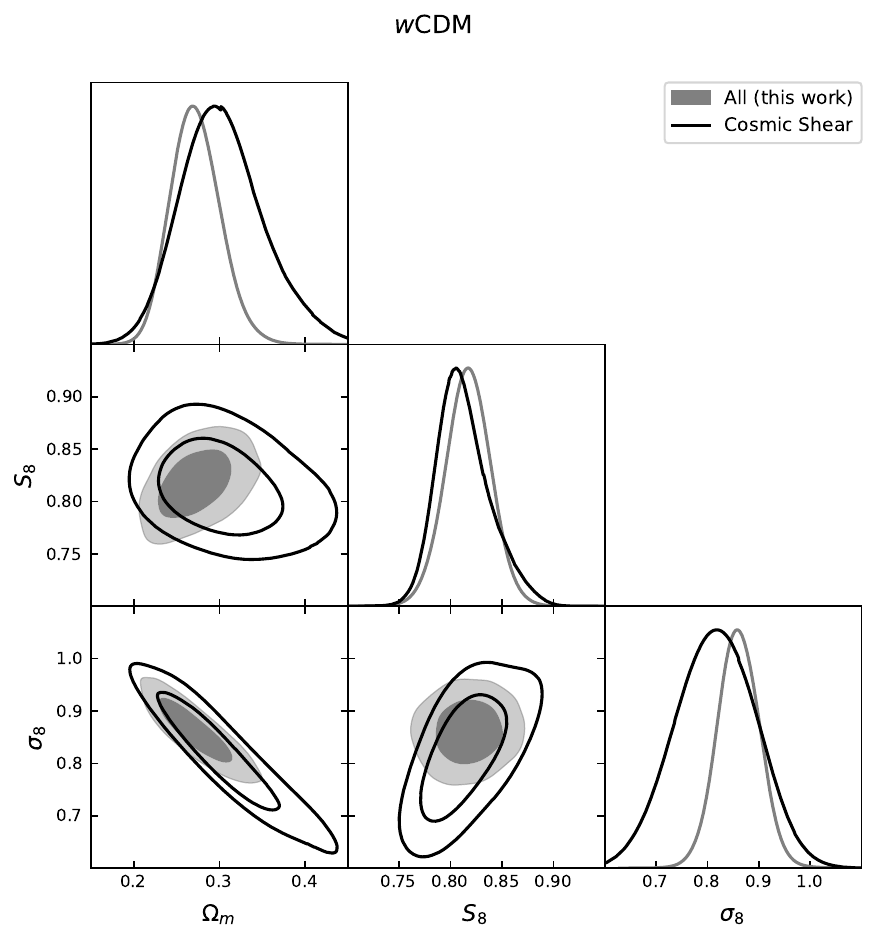}
\end{center}
\caption{Posterior distributions of the cosmological parameters $\Omega_{\textrm{m}}$, $S_8$, and $\sigma_8$,  as measured in data. We compare the results of this work to other analyses using DES data. The two-dimensional marginalised contours in these figures show the 68 percent and 95 percent credible regions.}
\label{fig:all_comparison}
\end{figure*}

\begin{figure*}
\begin{center}
\includegraphics[width=0.45\textwidth]{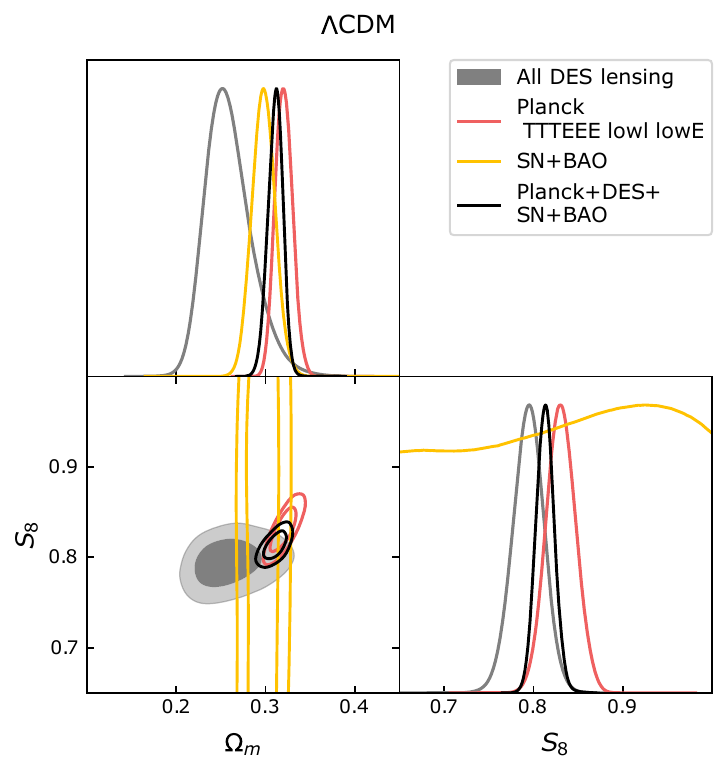}
\includegraphics[width=0.45\textwidth]{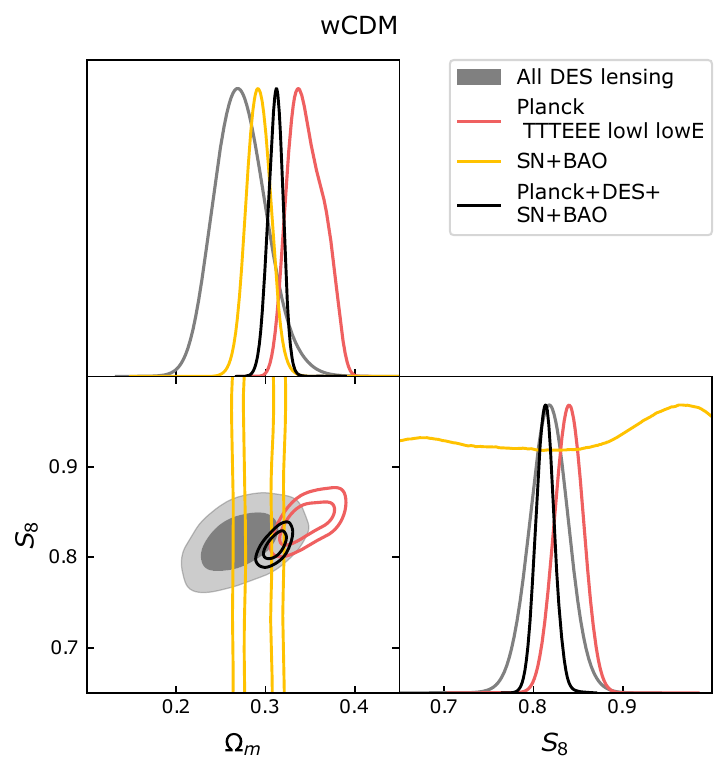}
\end{center}
    \caption{DES Y3 and external data constraints from low and high redshift probes. The two-dimensional marginalised contours in these figures show the 68 percent and 95 percent credible regions. We find no evidence for statistical inconsistency between the results of this work and external data set. }
\label{fig:all_highandlow}
\end{figure*}

We present the cosmological constraints obtained from our analysis of the DES Y3 data, assuming in turn the $\Lambda$CDM and $w$CDM models. We focus on the parameters used to learn the likelihood, which are the most constrained ones: $\Omega_{\textrm{m}}$, $S_8$, $A_{\textrm{IA}}$, and (for the $w$CDM scenario) $w$. The other cosmological parameters and nuisance parameters varied when creating the mocks are automatically marginalized over in our formalism. When reporting constraints, we also report the Figure of Merit (FoM) associated with $S_8$ and $\Omega_{\textrm{m}}$, defined by
\begin{equation}
{\textrm{FoM}(S_8,\Omega_{\textrm{m}}}) = \left({\textrm{det}} (C_{S_8,\Omega_{\textrm{m}}})\right)^{-0.5}.
\end{equation}
where $C_{S_8,\Omega_{\textrm{m}}}$ is the covariance matrix of the parameters (estimated from their posterior).

\subsection{$\Lambda$CDM results}
This subsection presents the $\Lambda$CDM results. The left plot in Fig. \ref{fig:all} displays the posterior distributions for $S_8$, $\Omega_{\textrm{m}}$, and $\sigma_8$ from a combination of different summary statistics; the constraints are also detailed in Table \ref{final_results_lcdm}, together with $p$-values and the FoMs. Constraints for the individual statistics are shown in Appendix \ref{Appendix_individual}. As confirmed by our PPD consistency tests (Table \ref{table:PPD}), all the posteriors appear to be consistent with each other, with their constraints largely overlapping. The marginalised mean values of $S_8$, $\Omega_{\textrm{m}}$, and $\sigma_8$ for the combination of all the summary statistics (grey line in plot), along with the 68 percent credible intervals, are:
\begin{eqnarray*}
S_8 = 0.794 \pm 0.017\\
\Omega_{\textrm{m}} = 0.259 \pm 0.025\\
\sigma_8 = 0.857 \pm  0.042 \, .\\
\end{eqnarray*}
%
%\LW{I'm surprised that the contours are symmetric about the mean; is this really the case? The 1D marginals don't look symmetric.} 
The constraints obtained by combining all the summary statistics are significantly tighter than the constraints from second moments only: by $\sim15 $ percent for $S_8$, $\sim 25$ percent for $\sigma_8$ and $\Omega_{\textrm{m}}$, and $\sim$70 percent for the FoM. Importantly, constraints significantly improve not only $S_8$ but also $\Omega_{\text{m}}$. When combining non-Gaussian statistics with second moments, WPH provides the largest gain, followed by ST and 3rd moments; the combination of all the summary statistics provide a further enhancement in the constraints, confirming the findings of \paperA{}. The gain in constraining power comes both from the additional information probed by non-Gaussian statistics and from the degeneracy breaking in the $\Omega_{\textrm{m}} \text{--} \sigma_8$ plane (as non-Gaussian statistics are characterized by a slightly different tilt in that plane compared to second moments).

Fig. \ref{fig:all_IA} shows the constraints on the intrinsic alignment parameter $A_{\textrm{IA}}$, for second moments and the combination of all summary statistics. The constraints improve by up to approximately 35 percent when non-Gaussian statistics are included, underscoring the ability of beyond-Gaussian statistics to enhance non-cosmological parameters as well. Our measurements indicate a preference for a small positive (i.e. non-zero) intrinsic alignment amplitude. 

\subsection{$w$CDM results}

This subsection analyzes the $w$CDM results. The right plot in Fig. \ref{fig:all} displays the posterior distributions for $S_8$, $\Omega_{\textrm{m}}$, and $\sigma_{8}$ for a combination of summary statistics (the same as shown for the $\Lambda$CDM case); constraints are also detailed in Table \ref{final_results_wcdm}, along with $p$-values and the Figures of Merit (FoMs). Fig. \ref{fig:all_w} shows the posterior for $S_8$ and $w$.
%\LW{Why not just show a single larger corner plot with all four $w$CDM parameters? Certainly I found it surprising that this wasn't shown, so maybe the reason for not doing so should be stated.} 
The results are in spirit very similar to what we obtained for the $\Lambda$CDM case; posteriors derived from different summary statistics largely overlap, and the combination of these statistics significantly improves constraints compared to those from second moments only. The marginalised mean values of $S_8$, $\Omega_{\textrm{m}}$, and $\sigma_8$ for the combination of all the summary statistics (grey line in plot), along with the 68 percent credible intervals, are:
\begin{eqnarray*}
S_8 = 0.817 \pm 0.021\\
\Omega_{\textrm{m}} = 0.273 \pm 0.029\\
\sigma_8 = 0.861 \pm  0.041 \, .\\
\end{eqnarray*}
%
%\LW{Again, are the contours really symmetric?}

We observe an improvement in the constraints over second moments of approximately 25 percent for $S_8$ and $\sigma_{8}$, approximately 20 percent for $\Omega_{\textrm{m}}$, and about 90 percent for the Figure of Merit (FoM) when all the statistics are considered. Interestingly, we obtain slightly higher values for $S_8$ in the $w$CDM scenario compared to the $\Lambda$CDM case. This is a consequence of opening the parameter space along $w$ in a non-symmetric way ($w > -1$); since the $S_8$ and $w$ posteriors show a slight degeneracy (see Fig. \ref{fig:all_w}), this pushes $S_8$ upwards compared to the case when $w$ is fixed at $-1$.  Concerning the constraints on the parameter $w$, all the probes are consistent with each other. The constraints get smaller as we include more summary statistics, and they push towards our prior boundary at $w=-1$, consistent with a $\Lambda$CDM scenario. Last, the constraints on the intrinsic alignment amplitude $A_{\textrm{IA}}$ are completely consistent with the ones obtained in the $\Lambda$CDM case, as shown in Fig. \ref{fig:all_IA}.

\subsection{Comparison with other DES analyses}\label{sect:des_comparison}

We discuss how the parameter constraints obtained from this work compare to those obtained by other cosmological analyses using DES Y3 weak lensing data. Several analyses incorporating Gaussian and non-Gaussian statistics have been conducted with DES Y3 data. Although these studies use identical data and calibration, direct comparison of their outcomes is complex. This complexity arises from varying analysis and modelling choices adopted in each study, as these can significantly impact the results. To facilitate a more accurate comparison, we attempted to replicate the analyses using similar analysis and modelling choices as in our study, wherever feasible. Below, we provide an overview of each study we are comparing. Fig. \ref{fig:all_comparison} offers a visual comparison of the constraints on $S_8$, $\sigma_{8}$, and $\Omega_{\textrm{m}}$ for both the $\Lambda$CDM and $w$CDM models, and the corresponding constraint values and Figures of Merit (FoMs) are detailed in Tables \ref{final_results_lcdm} and \ref{final_results_wcdm}.

\paragraph*{DES Y3 cosmic shear.} The first study to which we compare our results are the DES Y3 cosmic shear analyses \citep{y3-cosmicshear2,y3-cosmicshear1} for both $\Lambda$CDM and $w$CDM models. We repeated these analyses trying to match the same analysis choices used in our study. Notably, we excluded the shear ratio likelihood \citep{y3-shearratio}, since it is not included in our pipeline. Shear ratios, which are galaxy-galaxy lensing measurements at small scales, primarily constrain intrinsic alignment (IA) and redshift parameters. Since IA amplitude is correlated with $S_8$, incorporating shear ratios significantly enhances $S_8$ constraints. We also adopted the same priors for parameters $n_{\textrm{s}}$, $\Omega_{\textrm{b}}$, $h_{100}$, and neutrino mass used in this work. For redshift uncertainties, we used the \textsc{HyperRank} method \citep{y3-hyperrank}, which was not the standard method in \citep{y3-cosmicshear2,y3-cosmicshear1}, but which is employed in our current analysis. We also used as intrinsic alignment model the Non-Linear Alignment (NLA) model, augmented with a clustering term with a fixed galaxy-matter bias of one.  We note, however, that the clustering term included in the IA model of the cosmic shear analysis is estimated using tree-level perturbation theory, whereas our implementation directly uses the clustering of the simulation, which should be more accurate. Another difference with our pipeline is that it automatically excludes information on large multipoles ($\ell>1000$) by imposing a hard cut during the map making process. This restriction is not present in the cosmic shear analysis, which can theoretically probe smaller scales. Both analyses, however, have undergone the same process of scale cuts, where small scales were excluded if they were potentially affected by unmodelled baryonic effects. Last, the full impact of source clustering is not accounted for in the cosmic shear analysis, although its impact should be negligible \citep{sourceclustering}.

Fig. \ref{fig:all_comparison} and Tables \ref{final_results_lcdm} and \ref{final_results_wcdm} show the remarkable agreement between the DES cosmic shear analysis and our work. The cosmic shear Figure of Merit (FoM) is very similar to that from our second moments; this is expected, as they are both Gaussian statistics applied to the same dataset with similar analysis choices. The FoM from our work, when all summary statistics are combined, greatly improves over the cosmic shear one, as expected. The constraints on intrinsic alignment and $w$ are also compatible. Interestingly, despite the FoM from the combination of all statistics being almost double that of the cosmic shear analysis, we note that not all parameter constraints are necessarily improved (e.g. $A_{\textrm{IA}}$ in the $w$CDM case or $S_8$ in the $\Lambda$CDM case). This might be due to the limited access to small-scale information at $\ell > 1000$ not probed by our work, a non-optimal compression of our statistics along the $S_8$ and $A_{\textrm{IA}}$ directions, or simply due to a slightly different degeneracy direction in the $\sigma_{8} \text{--} \Omega_{\textrm{m}}$ plane, a consequence of different constraining power. Despite this, the significant gain in the FoM over the fiducial DES cosmic shear analysis, and the compatibility of the constraints obtained using a completely different and independent pipeline, strongly highlight the importance of a simulation-based analysis using non-Gaussian statistics.

\paragraph*{DES Y3 second and third moments.} The second study to which we compare our results is the $\Lambda$CDM analysis of second and third moments from Gatti et al. \cite{moments2021}. This analysis uses the third moments as non-Gaussian statistics, measured in the same way as we did in this work. However, the modelling approach was analytical, in contrast to our pipeline. Additionally, a MOPED compression \citep{Heavens2000} was used, in contrast to the neural network compression used in the current study. As with the case of cosmic shear, we endeavoured to align the analysis choices with our pipeline. In contrast to the fiducial analysis choices in \cite{moments2021}, we reran the second and third moments analysis without including shear ratios; furthermore, we matched the priors on the parameters $n_{\textrm{s}}$, $\Omega_{\textrm{b}}$, and $h_{100}$. However, the \cite{moments2021} analysis did not vary the neutrino mass, as neutrinos were not modelled; this was an analysis choice we could not alter. Additionally, source clustering effects were not modelled, although their impact on third moments was mitigated using the method explained in \cite{sourceclustering}. For the intrinsic alignment model, we employed the NLA model, but we were unable to include a galaxy-matter bias term. Lastly, as in the cosmic shear analysis, no explicit cut on multipoles $\ell > 1000$ was made, though a scale cut intended to minimize the impact of unmodelled baryonic effects was implemented.

Results from this revised analysis of the second and third moments also overlap with the constraints from our work (see Fig. \ref{fig:all_comparison}, and Tables \ref{final_results_lcdm} and \ref{final_results_wcdm}). The Figure of Merit (FoM) is larger than that of our second moments, but smaller than the one from the combination of all summary statistics. The constraints are also compatible with those obtained from our implementation of the second and third moments analysis. Given the subtle differences between the two pipelines, we did not expect a perfect match; however, the match is notably good. This serves as further validation of the work done here and the analytical modelling in \cite{moments2021}, considering that the two pipelines are completely independent and follow very different approaches.

\paragraph*{DES Y3 Peak counts.} The third study to which we compare our results are the $\Lambda$CDM constraints from the power spectrum plus peak counts analysis presented in Zuercher et al. \cite{Zuercher2021b}. This analysis includes peak counts as a non-Gaussian statistic and is entirely simulation-based, though it differs from our work in several aspects. 

Firstly, the analysis relied solely on the \texttt{DarkGridV1} simulation suite, primarily exploring two cosmological parameters ($S_8$ and $\Omega_{\textrm{m}}$). The dependence on other cosmological parameters ($n_{\textrm{s}}$, $\Omega_{\textrm{b}}$, and $h_{100}$) was modelled through a Taylor expansion of the data vector, calibrated against a limited set of N-body simulations. Neutrinos were not varied in this analysis. To model the data vector, a three-parameter emulator was constructed for $S_8$, $\Omega_{\textrm{m}}$, and $A_{\textrm{IA}}$; other nuisance parameters were modelled by Taylor-expanding the data vector, an approach different from ours. A Non-Linear Alignment (NLA) intrinsic alignment (IA) model was used, without the clustering term, and source clustering effects were not included. As in our work, no shear ratio likelihood was incorporated in the original analysis. No explicit cut on multipoles $\ell > 1000$ was made, although a scale cut was implemented to reduce the impact of unmodelled baryonic effects. A MOPED compression was used to compress the data vector, and a Gaussian likelihood was assumed. In our comparison with the peaks analysis, we retained all the fiducial choices made by \cite{Zuercher2021b}. We refrained from imposing the Gower Street prior on $n_{\textrm{s}}$, $\Omega_{\textrm{b}}$, and $h_{100}$, as this would have tightened the constraints and invalidated the original scale cut. For the cosmic shear and second and third moments analyses, such an adjustment was feasible as we chose to not include shear ratios, which counterbalanced the gain in constraining power from having a tight prior on $n_{\textrm{s}}$, $\Omega_{\textrm{b}}$, and $h_{100}$. 

The results of the peak counts analysis, as displayed in Fig. \ref{fig:all_comparison}, are in alignment with our findings, with the posterior largely overlapping ours. Notably, this analysis delivers tighter constraints on $S_8$ but is much less restrictive regarding $\Omega_{\textrm{m}}$ and the overall Figure of Merit. The noticeable differences in constraining power, given the significant differences in analysis choices, are somewhat expected. This underscores the challenges in comparing results from the same dataset when different analysis methods are employed.

\section{Comparison with external data}\label{sec:external_data}

We compare our cosmological constraints with results from various external datasets. Although different choices could be made regarding which data to compare, we have adhered to the comparisons used in the fiducial DES cosmic shear analysis \citep*{y3-cosmicshear2}. In particular we consider:
\begin{itemize}
    \item{
        \textbf{eBOSS}: we use spectroscopic baryon acoustic oscillation (BAO) measurements from eBOSS \citep{allam20}. 
        The BAO likelihood is assumed to be independent of DES. See \cite{y3-3x2ptkp} for a detailed description of the likelihoods included (but note that unlike \cite{y3-3x2ptkp}, we do not include RSD constraints from eBOSS; this is in line with the analysis choices adopted for the DES Y3 cosmic shear analysis in \citep*{y3-cosmicshear2}).
    }
    \item{
        \textbf{Pantheon Supernovae}: we consider the luminosity distances from Type Ia supernovae from the Pantheon sample \citep{scolnic18}; this data set includes 279 type Ia supernovae from the PanSTARRS Medium Deep Survey $(0.03 < z < 0.68)$ and samples from SDSS, SNLS, and HST. The final Pantheon catalogue includes 1048 SNe, out to $z = 2.26$.
    }
    \item{
        \textbf{\planck{} 2018 CMB}: we use the final data release of the \planck{} Cosmic Microwave Background (CMB) survey \citep{aghanim19} including measurements of both temperature and polarization anisotropies, in the same way it is implemented in \citep*{y3-cosmicshear2}.
    }
\end{itemize}
To perform a meaningful comparison, we recomputed the posteriors using the priors on cosmological parameters used in this work (Table \ref{parameter}) for eBOSS and supernovae data. We did not do this for the \planck{} data as the priors on $n_{\rm s}$, $h$, and $\Omega_{\rm b}$ are partially determined by \planck{} constraints. Note, though, that we are mostly interested in testing the compatibility for the parameters $S_8$, $\Omega_{\textrm{m}}$, and $w$; our analysis does not constrain the other cosmological parameters, which are completely prior dominated.

\subsection{Quantifying tension}
To estimate the agreement or disagreement between different datasets, we use the normalizing flow Monte Carlo estimate of the probability of a parameter difference as discussed in \citep{Raveri:2021wfz}. We use the implementation of this algorithm in \texttt{tensiometer}~\citep{Raveri:2021wfz, Raveri:2018wln}. 
In the case of uncorrelated data sets, the probability of the parameter difference is:
\begin{equation} \label{Eq:ParameterDifferencePDF}
\mathcal{P}(\Delta \theta) = \int_{V_p} \mathcal{P}_A(\theta) \mathcal{P}_B(\theta-\Delta \theta) d\theta,
\end{equation}
where $V_p$ is the prior support and $\mathcal{P}_A$ and $\mathcal{P}_B$ are the two posterior distributions of the parameters. 
The probability of an actual shift in parameter space is obtained from the density of parameter shifts:
\begin{equation} \label{Eq:ParamShiftProbability}
\Delta = \int_{\mathcal{P}(\Delta\theta)>\mathcal{P}(0)} \mathcal{P}(\Delta\theta) \, d\Delta\theta,
\end{equation}
which is the posterior mass above the contour of constant probability for no shift, $\Delta\theta=0$. 
Due to the discrete nature of our posterior samples, the convolution integral in Eq.~\eqref{Eq:ParameterDifferencePDF} is performed with a Monte Carlo algorithm. 
To compute Eq.~\eqref{Eq:ParamShiftProbability} we first train a normalizing flow within \texttt{tensiometer} to model the probability density of parameter shifts and then we perform a Monte Carlo integral with the trained normalizing flow. For further details, we refer the reader to \citep{Raveri:2021wfz}.\\

In the following, our results are always reported in terms of effective
number of standard deviations. When considering an event with probability $\mathcal{P}$, this is determined by
\begin{equation} \label{Eq:NumberSigma}
n_{\sigma} \equiv \sqrt{2} \mathrm{Erf}^{-1}(\mathcal{P}),
\end{equation}
where $\mathrm{Erf}^{-1}$ is the inverse error function.
It represents the number of standard deviations that an event with the same probability would have if it were drawn from a Gaussian distribution. This definition does not assume a Gaussian distribution of the underlying statistics and should be interpreted as a logarithmic scale for probabilities.

\subsubsection{Compatibility between this analysis and external data}

In Table~\ref{tab:external_tension_table} we show the estimate of the tension between the different DES observables we consider, the \planck{} 2018 CMB data, and the joint BAO+SNe dataset. As we can see DES measurements are broadly in agreement with these measurements.
In particular, we find that no pairwise comparison exceeds $2.2\sigma$. 
This tells us that different DES data combinations agree with geometric probes on the determination of $\Omega_m$ and $w$ and with the CMB on all three parameters. 

\begin{table}[H]
    \centering
    \begin{tabular}{l c c}
    \toprule
    \textbf{$\Lambda$CDM Summary Statistic(s)} &  \textbf{\planck{} 2018} &  \textbf{BAO+SNe}  \\
    \midrule
      2nd moments  & 1.4$\sigma$ &0.7$\sigma$\\
      2nd+3rd moments  & 1.4$\sigma$ & 0.8$\sigma$ \\
      2nd moments+ST  & 1.7$\sigma$ & 0.9$\sigma$ \\
      2nd moments+WPH   & 1.2$\sigma$ &0.8$\sigma$ \\
      2nd moments+ST+WPH   & 2.0$\sigma$ & 1.3$\sigma$ \\
      2nd+3rd moments+ST+WPH & 1.8$\sigma$ & 1.1$\sigma$ \\  
    \bottomrule
    \\
    \toprule
    \textbf{$w$CDM Summary Statistic(s)} & \textbf{\planck{} 2018} & \textbf{BAO+SNe}  \\
    \midrule
      2nd moments  & 1.8$\sigma$ & 0.9$\sigma$ \\
      2nd+3rd moments  & 1.6$\sigma$ & 0.7$\sigma$ \\
      2nd moments+ST  & 2.0$\sigma$ & 0.8$\sigma$ \\
      2nd moments+WPH   & 1.8$\sigma$ & 1.0$\sigma$ \\
      2nd moments+ST+WPH   & 1.6$\sigma$ & 1.0$\sigma$ \\
      2nd+3rd moments+ST+WPH & 2.2$\sigma$ & 1.0$\sigma$ \\  
    \bottomrule
    \end{tabular}
    \caption{ \label{tab:external_tension_table}
    Compatibility of different DES observables with \planck{} and BAO + SNe results, assuming the $\Lambda$CDM and $w$CDM models.
    }
\end{table}

In Table~\ref{tab:LCDM_tension_joint_external_table} we show the results of estimating the discrepancy between DES probes and all other joint datasets, assuming both the $\Lambda$CDM and $w$CDM models.
This reassures us that all DES data combinations can be combined with all external probes. Note that (in more than one dimension) having $A$ and $B$ in agreement, and having both separately in agreement with $C$, does not guarantee that the combination of $A$ and $B$ will be in agreement with $C$.
For this reason, we explicitly check for the compatibility of all DES data combinations and joint external data.

\begin{table}[H]
    \centering
    \begin{tabular}{l c}
    \toprule
    \textbf{$\Lambda$CDM Summary Statistic(s)} & \textbf{\planck{} 2018+BAO+SNe}  \\
    \midrule
      2nd moments  & 1.1$\sigma$ \\
      2nd+3rd moments  & 1.1$\sigma$  \\
      2nd moments+ST  & 1.4$\sigma$  \\
      2nd moments+WPH   & 1.1$\sigma$  \\
      2nd moments+ST+WPH   & 1.7$\sigma$ \\
      2nd+3rd moments+ST+WPH & 1.6$\sigma$ \\  
    \bottomrule
    \\
    \toprule
    \textbf{$w$CDM Summary Statistic(s)} & \textbf{\planck{} 2018+BAO+SNe}  \\
    \midrule
      2nd moments  & 1.4$\sigma$ \\
      2nd+3rd moments  & 1.2$\sigma$  \\
      2nd moments+ST  & 1.6$\sigma$  \\
      2nd moments+WPH   & 1.5$\sigma$  \\
      2nd moments+ST+WPH   & 1.6$\sigma$ \\
      2nd+3rd moments+ST+WPH & 1.7$\sigma$ \\  
    \bottomrule
    \end{tabular}
    \caption{ \label{tab:LCDM_tension_joint_external_table}
    Compatibility of different DES observables with \planck{} + BAO + SNe results, assuming the $\Lambda$CDM and $w$CDM models.
    }    
\end{table}

\subsubsection{Compatibility between early and late times universe}

We now investigate the agreement between probes of the early and late time universe.
We select the DES probe combining  `2nd+3rd moments+ST+WPH' since it is the most constraining among the DES summary statistics.
Table~\ref{tab:early_late_tension_table} shows broad agreement between these two probes for both the $\Lambda$CDM and $w$CDM models.

\begin{table}[H]
    \centering
    \begin{tabular}{c c}
    \toprule
    &   \textbf{Late (DES+SNe+BAO)} vs \textbf{Early (\planck{} 2018)}    \\
    \midrule
       $\Lambda$CDM & 1.7$\sigma$ \\
      $w$CDM & 2.1$\sigma$ \\  
    \bottomrule
    \end{tabular}
    \caption{ \label{tab:early_late_tension_table}
    Compatibility of probes of the early and late time universe within the $\Lambda$CDM and $w$CDM models.
    For the early Universe, we consider CMB measurements from \planck{} 2018. For the late-time Universe we use the 2nd+3rd moments+ST+WPH combination of DES probes joined with SNe and BAO measurements.
    }
\end{table}

\subsubsection{Joint constraints}

Having found a good level of agreement between the posterior probability distributions from this analysis and those from external probes, we can be confident that they can be combined.

As the external data and our DES Y3 observations are independent, we can simply use the posterior probability distributions from the external data as the prior probability for our analysis.

We have already used the \planck{} external data to justify the prior probability distributions for the so-called nuisance parameters in the Gower Street simulations (e.g. for $\Omega_b$). Avoiding double counting of information, we simply use a new prior distribution only on the parameters of interest (i.e. $S_8$, $\Omega_m$, and $w$), which is given by the posterior distribution from the external analyses. As our original priors on the parameters of interest are flat, this requires a simple reweighting of the existing posterior distribution.

We do this in two ways: (i) relearning the posterior density of the parameters of interest from both our new DES Y3 MCMC chains and the MCMC chains from the external probes – we then multiply the learned densities; (ii) learning the posterior density of the parameters of interest for the external probes, and rerunning the MCMC chains using our learned DES Y3 simulation-based likelihood using the learned density as a prior. In both cases we use normalizing flows implemented in the package \texttt{tensiometer} to learn the densities from the existing posterior samples.

Both these methods give consistent results; we use method (i) in all quoted results.

%\NJ{Marco, you now have to do this second method test! :) }

% Having found a good level of agreement between all probes we consider we can proceed with combining their constraints.
% To do so we join and sample the normalizing flows obtained with \texttt{tensiometer}, trained on the single posteriors.
% The normalizing flows in \texttt{tensiometer} are optimized, as discussed in~\citep{Raveri:2021wfz} to accurately reproduce the posteriors we consider and these posteriors overlap significantly -- as measured by a lack of tension between them. Hence we expect the joint results to be reliable. 

\begin{table}[H]
    \centering
    \begin{tabular} { l c c c}
    \toprule
    \textbf{$\Lambda$CDM} &  \textbf{DES+SNe+BAO} & \textbf{\planck{} 2018} & \textbf{Joint}\\
    \midrule
    {$S_8$            } & $0.800\pm 0.017            $ & $0.831\pm 0.016             $ & $0.810\pm 0.011           $\\
    {$\Omega_m       $} & $0.290^{+0.010}_{-0.012} $ & $0.3206\pm 0.0096 $         & $0.3070\pm 0.0067 $\\
    \bottomrule
    \\
    \toprule
    \textbf{$w$CDM} &  \textbf{DES+SNe+BAO} & \textbf{\planck{} 2018} & \textbf{Joint}\\
    \midrule
    {$S_8            $} & $0.807\pm 0.020           $ & $0.840\pm 0.016             $ & $0.813\pm 0.010            $\\
    {$\Omega_m       $} & $0.285\pm 0.013       $& $0.346^{+0.017}_{-0.024}   $ & $0.3106\pm 0.0085          $\\
    {$w          $} & $< -0.905$ & $< -0.891$ & $< -0.97$\\
    \bottomrule
    \end{tabular}
    \caption{ \label{tab:joint.constraints}
    Cosmological constraints on parameters of interest in the $\Lambda$CDM and $w$CDM models. For each parameter we report the mean and 68 percent credible interval.
    }
\end{table}

As we can see from Table~\ref{tab:joint.constraints} the constraining power from late-time universe probes is comparable with that of CMB measurements. In $\Lambda$CDM this roughly amounts to a 2\% constraint on $S_8$ and a 3\% constraint on $\Omega_m$. Since the two probes are independent the constraints add in quadrature almost perfectly, resulting in a 1\% level joint constraint. The joint constraints are also shown in Fig. \ref{fig:all_highandlow}.

\section{Summary}\label{sect:summary}

We presented a cosmological analysis using a combination of Gaussian and non-Gaussian statistics applied to the weak lensing mass (convergence) maps from the first three years (Y3) of the Dark Energy Survey (DES). Specifically, we considered: 1) second and third moments; 2) wavelet phase harmonics (WPH); 3) the scattering transform (ST). The second moments are Gaussian statistics, while the third moments probe additional non-Gaussian information of the fields. The WPH moments are the second moments of smoothed weak lensing mass maps that have undergone a non-linear transformation, enabling the exploration of the non-Gaussian features of the field. The ST coefficients are generated through a series of smoothing and modulus operations applied to the input field, followed by averaging. Both the WPH and ST are often linked to convolutional neural networks (CNNs) because their statistical definitions bear similarities to the architecture of CNNs. However, unlike CNNs, they do not require training data.

Our analysis is entirely based on simulations, spanning a space of seven $\nu w$CDM cosmological parameters. It forward-models the most relevant sources of systematic errors in the data, including masks, noise variations, clustering of the sources, intrinsic alignments, and shear and redshift calibration. We have implemented a neural network to compress the summary statistics, and estimated the parameter posteriors using a simulation-based inference (SBI) approach. Our analysis setup has been extensively validated in \paperA{}, which includes the validation of the mocks used in the analysis, empirical coverage tests of the posteriors obtained via simulation-based inference, tests on the impact of physical effects not included in the simulations (e.g. baryonic feedback effect), and an end-to-end validation of the pipeline using an independent set of simulations. We complemented the tests from \paperA{} with additional ones, particularly focusing on our data measurements. These include tests on the internal consistency of the different summary statistics measured in the data, assessments of goodness-of-fit, and B-mode null tests. 

We demonstrated that incorporating and integrating various non-Gaussian statistics significantly enhances constraints compared to relying solely on Gaussian statistics (specifically, second moments). Notably, we achieved a 70 percent (90 percent) improvement in the Figure of Merit $\textrm{FoM}(S_8, \Omega_{\textrm{m}})$ for the $\Lambda$CDM ($w$CDM) model. The improvement in the \(\textrm{FoM}(S_8, \Omega_{\textrm{m}})\) is a result of significant enhancements in both \(S_8\) and \(\Omega_{\textrm{m}}\). By combining all summary statistics, we measured the amplitude of fluctuation parameter $S_8 \equiv \sigma_8 (\Omega_{\textrm{m}}/0.3)^{0.5} = 0.794 \pm 0.017$ with 2 percent precision, under the assumption of a $\Lambda$CDM cosmology. In the case of a $w$CDM cosmology, the measurement was $S_8 = 0.817 \pm 0.021$. The inclusion of non-Gaussian statistics significantly tightened the constraints on $A_{\textrm{IA}}$, the amplitude of intrinsic alignment. Additionally, in the context of the $w$CDM scenario, these statistics also strengthened the constraints on the parameter $w$, obtaining $w<-0.72$, consistent with a $\Lambda$CDM scenario.

We compared our results with other weak lensing results from the DES Y3 data, finding good consistency. Our constraints outperform the results from the fiducial DES cosmic shear analysis (as expected) due to the extra information probed by the non-Gaussian statistics. We also find statistical agreement ($<2.2\sigma$) when comparing with results from external datasets: \planck{} constraints from the Cosmic Microwave Background, constraints from spectroscopic BAO measurements, and constraints from Type Ia supernovae.

%Last, A recently published companion DES analysis (\citealt{jeffrey2024dark}) used the same simulation-based inference pipeline as described in this paper, but using convolutional neural networks (map-level inference), peak counts, and power spectra to infer cosmology. Their results are consistent with the results presented in this work. All of these are analyses of DES Y3 data, pending the final full DES Year 6 data.
%We expect to improve the analysis presented in this work and apply it to future data, such at the final DES Y6 data... 

%Based on the investigation performed in \G\, we expect to further improve our constraining power on $S_8$ by roughly 20 percent, if we take into account the expected increase in the source number density. We plan to be able to model baryonic effects, which should allow us to push our analysis to smaller scales, improving constraints (up to 20 percent, Appendix \ref{sect:scale_cuts}) and learning about baryonic physics. We are also planning to expand our modelling to include massive neutrinos and the full $w$CDM parameter space.

\section*{Contribution Statement}
%MG: Designed the project, led the analysis (created the mocks, performed the measurements, tested the pipeline, and obtained cosmological constraints), interpreted the results, and led the writing of the manuscript.
%This paper has undergone DES internal review; DA, SS and TK were the DES internal reviewer. 
MG: Designed the project, led the analysis, and led the writing of the manuscript.
GC: Performed comparisons with external datasets and edited the manuscript.
NJ: Co-led the generation of the Gower Street simulations, proposed the simulation-based inference workflow, assisted with result interpretation, and edited the manuscript.
LW: Co-led the generation of the Gower Street simulations, suggested mathematical notation, and edited the manuscript.
AP: Recomputed DES Y3 cosmic shear posteriors using Gower Street priors.
JP: Recomputed external dataset posteriors using Gower Street priors.
JW: Helped perform measurements in simulations and helped with mocks testing.
MR: Helped with interpretation of tension metrics, and with editing of the manuscript.
BJ: Assisted with the interpretation of results and edited the manuscript.
VA, GG, MY, CZ: Helped perform measurements in simulations.
JB: Provided feedback on analyses choices and intrinsic alignment implementation.
DA, SS, TK: DES internal reviewers of the paper.
MG, ES, AA, MB, MT, AC, CD, NM, ANA, IH, DG, GB, MJ, LFS, AF, JM, RPR, RC, CC, SP, IT, JP, JEP, CS: Production of the DES shape catalog.
JM, AA, AA, CS, SE, JD, JM, DG, GB, MT, SD, AC, NM, BY, MR: Production of the DES redshift distribution.
SE,BY, NK, EH, YZ: Production of the DES Balrog.
MJ, GB, AA, CD, PFL, KB, IH, MG, AR: Production of the DES PSF.
Builders: The remaining authors have made contributions to this paper that include, but are not limited to, the construction of DECam and other aspects of collecting the data; data processing and calibration; developing broadly used methods, codes, and simulations; running the pipelines and validation tests; and promoting the science analysis.

\section*{Acknowledgements}

The Gower Street simulations were generated under the DiRAC project p153 `Likelihood-free inference with the Dark Energy Survey' (ACSP255/ACSC1) using DiRAC (STFC) HPC facilities (\url{www.dirac.ac.uk}).

Funding for the DES Projects has been provided by the U.S. Department of Energy, the U.S. National Science Foundation, the Ministry of Science and Education of Spain, 
the Science and Technology Facilities Council of the United Kingdom, the Higher Education Funding Council for England, the National Center for Supercomputing 
Applications at the University of Illinois at Urbana-Champaign, the Kavli Institute of Cosmological Physics at the University of Chicago, 
the Center for Cosmology and Astro-Particle Physics at the Ohio State University,
the Mitchell Institute for Fundamental Physics and Astronomy at Texas A\&M University, Financiadora de Estudos e Projetos, 
Funda{\c c}{\~a}o Carlos Chagas Filho de Amparo {\`a} Pesquisa do Estado do Rio de Janeiro, Conselho Nacional de Desenvolvimento Cient{\'i}fico e Tecnol{\'o}gico and 
the Minist{\'e}rio da Ci{\^e}ncia, Tecnologia e Inova{\c c}{\~a}o, the Deutsche Forschungsgemeinschaft and the Collaborating Institutions in the Dark Energy Survey. 

The Collaborating Institutions are Argonne National Laboratory, the University of California at Santa Cruz, the University of Cambridge, Centro de Investigaciones Energ{\'e}ticas, 
Medioambientales y Tecnol{\'o}gicas-Madrid, the University of Chicago, University College London, the DES-Brazil Consortium, the University of Edinburgh, 
the Eidgen{\"o}ssische Technische Hochschule (ETH) Z{\"u}rich, 
Fermi National Accelerator Laboratory, the University of Illinois at Urbana-Champaign, the Institut de Ci{\`e}ncies de l'Espai (IEEC/CSIC), 
the Institut de F{\'i}sica d'Altes Energies, Lawrence Berkeley National Laboratory, the Ludwig-Maximilians Universit{\"a}t M{\"u}nchen and the associated Excellence Cluster Universe, 
the University of Michigan, NSF's NOIRLab, the University of Nottingham, the Ohio State University, the University of Pennsylvania, the University of Portsmouth, 
SLAC National Accelerator Laboratory, Stanford University, the University of Sussex, Texas A\&M University, and the OzDES Membership Consortium.

Based in part on observations at Cerro Tololo Inter-American Observatory at NSF's NOIRLab (NOIRLab Prop. ID 2012B-0001; PI: J. Frieman), which is managed by the Association of Universities for Research in Astronomy (AURA) under a cooperative agreement with the National Science Foundation.

The DES data management system is supported by the National Science Foundation under Grant Numbers AST-1138766 and AST-1536171.
The DES participants from Spanish institutions are partially supported by MICINN under grants ESP2017-89838, PGC2018-094773, PGC2018-102021, SEV-2016-0588, SEV-2016-0597, and MDM-2015-0509, some of which include ERDF funds from the European Union. IFAE is partially funded by the CERCA program of the Generalitat de Catalunya.
Research leading to these results has received funding from the European Research
Council under the European Union's Seventh Framework Program (FP7/2007-2013) including ERC grant agreements 240672, 291329, and 306478.
We  acknowledge support from the Brazilian Instituto Nacional de Ci\^encia
e Tecnologia (INCT) do e-Universo (CNPq grant 465376/2014-2).

This manuscript has been authored by Fermi Research Alliance, LLC under Contract No. DE-AC02-07CH11359 with the U.S. Department of Energy, Office of Science, Office of High Energy Physics.

\bibliography{bibliography}

\begin{thebibliography}{89}
\providecommand{\natexlab}[1]{#1}
\providecommand{\url}[1]{\texttt{#1}}
\providecommand{\urlprefix}{URL }
\providecommand{\eprint}[1][]{\url{#1}}

\bibitem[{{Abbott} et~al.(2022){Abbott} \& {Aguena} et~al.}]{y3-3x2ptkp}
{Abbott}, T.~M.~C., {Aguena}, M., {Alarcon}, A., et~al., 2022, \prd, 105, 2, 023520, \eprint arXiv:{2105.13549}

\bibitem[{{Ajani} et~al.(2020){Ajani} \& {Peel} \& {Pettorino} \& {Starck} \& {Li} \& {Liu}}]{ajani_peaks}
{Ajani}, V., {Peel}, A., {Pettorino}, V., {Starck}, J.-L., {Li}, Z., {Liu}, J., 2020, \prd, 102, 10, 103531

\bibitem[{{Alam} et~al.(2021){Alam} \& {Aubert} et~al.}]{allam20}
{Alam}, S., {Aubert}, M., {Avila}, S., et~al., 2021, \prd, 103, 8, 083533, \eprint arXiv:{2007.08991}

\bibitem[{{Allys} et~al.(2020){Allys} \& {Marchand} \& {Cardoso} \& {Villaescusa-Navarro} \& {Ho} \& {Mallat}}]{Allys2020}
{Allys}, E., {Marchand}, T., {Cardoso}, J.~F., {Villaescusa-Navarro}, F., {Ho}, S., {Mallat}, S., 2020, \prd, 102, 10, 103506

\bibitem[{{Alsing} et~al.(2018){Alsing} \& {Wandelt} \& {Feeney}}]{Alsing2018}
{Alsing}, J., {Wandelt}, B., {Feeney}, S., 2018, \mnras, 477, 2874

\bibitem[{{Amon} et~al.(2022){Amon} \& {Gruen} et~al.}]{y3-cosmicshear1}
{Amon}, A., {Gruen}, D., {Troxel}, M.~A., et~al., 2022, \prd, 105, 2, 023514

\bibitem[{{Anbajagane} et~al.(2023){Anbajagane} \& {Chang} et~al.}]{Anbajagane2023}
{Anbajagane}, D., {Chang}, C., {Banerjee}, A., et~al., 2023, \mnras, 526, 4, 5530, \eprint arXiv:{2308.03863}

\bibitem[{{Aric{\`o}} et~al.(2020){Aric{\`o}} \& {Angulo} et~al.}]{Arico2020}
{Aric{\`o}}, G., {Angulo}, R.~E., {Hern{\'a}ndez-Monteagudo}, C., et~al., 2020, \mnras, 495, 4, 4800

\bibitem[{{Asgari} et~al.(2021){Asgari} \& {Lin} et~al.}]{Asgari_2021}
{Asgari}, M., {Lin}, C.-A., {Joachimi}, B., et~al., 2021, \aap, 645, A104, \eprint arXiv:{2007.15633}

\bibitem[{{Banerjee} \& {Abel}(2023)}]{Banerjee2023}
{Banerjee}, A., {Abel}, T., 2023, \mnras, 519, 4, 4856

\bibitem[{{Barthelemy} et~al.(2020){Barthelemy} \& {Codis} \& {Uhlemann} \& {Bernardeau} \& {Gavazzi}}]{Barthelemy2020}
{Barthelemy}, A., {Codis}, S., {Uhlemann}, C., {Bernardeau}, F., {Gavazzi}, R., 2020, \mnras, 492, 3, 3420

\bibitem[{Bishop(1994)}]{Bishop1994}
Bishop, C.~M., 1994, Mixture density networks, Tech. rep., Aston University

\bibitem[{{Boruah} et~al.(2022){Boruah} \& {Lavaux} \& {Hudson}}]{Boruah2022}
{Boruah}, S.~S., {Lavaux}, G., {Hudson}, M.~J., 2022, \mnras, 517, 3, 4529

\bibitem[{{Boyle} et~al.(2021){Boyle} \& {Uhlemann} et~al.}]{Boyle2021}
{Boyle}, A., {Uhlemann}, C., {Friedrich}, O., et~al., 2021, \mnras, 505, 2, 2886

\bibitem[{{Bridle} \& {King}(2007)}]{Bridle2007}
{Bridle}, S., {King}, L., 2007, New Journal of Physics, 9, 444

\bibitem[{{Bruna} \& {Mallat}(2013)}]{Bruna2013}
{Bruna}, J., {Mallat}, S., 2013, arXiv e-prints, arXiv:1311.0407

\bibitem[{{Chang} et~al.(2018){Chang} \& {Pujol} et~al.}]{Chang2018}
{Chang}, C., {Pujol}, A., {Mawdsley}, B., et~al., 2018, \mnras, 475, 3165

\bibitem[{{Cheng} et~al.(2024){Cheng} \& {Marques} et~al.}]{Cheng2024}
{Cheng}, S., {Marques}, G.~A., {Grand{\'o}n}, D., et~al., 2024, arXiv e-prints, arXiv:2404.16085, \eprint arXiv:{2404.16085}

\bibitem[{{Cheng} et~al.(2020){Cheng} \& {Ting} \& {M{\'e}nard} \& {Bruna}}]{Cheng2020}
{Cheng}, S., {Ting}, Y.-S., {M{\'e}nard}, B., {Bruna}, J., 2020, \mnras, 499, 4, 5902

\bibitem[{{Cohen} \& {Ryan}(1995)}]{Cohen1995}
{Cohen}, A., {Ryan}, R.~D., 1995, Wavelets and multiscale signal processing, Chapman \& Hall

\bibitem[{{Cordero} et~al.(2022){Cordero} \& {Harrison} et~al.}]{y3-hyperrank}
{Cordero}, J.~P., {Harrison}, I., {Rollins}, R.~P., et~al., 2022, \mnras, 511, 2, 2170

\bibitem[{Dalal et~al.(2023)Dalal \& Li et~al.}]{HSC1}
Dalal, R., Li, X., Nicola, A., et~al., 2023, Phys. Rev. D, 108, 123519

\bibitem[{{Dietrich} \& {Hartlap}(2010)}]{Dietrich2010}
{Dietrich}, J.~P., {Hartlap}, J., 2010, \mnras, 402, 2, 1049

\bibitem[{{Doux} et~al.(2021){Doux} \& {Baxter} et~al.}]{Doux2021}
{Doux}, C., {Baxter}, E., {Lemos}, P., et~al., 2021, \mnras, 503, 2, 2688

\bibitem[{{Feldbrugge} et~al.(2019){Feldbrugge} \& {van Engelen} \& {van de Weygaert} \& {Pranav} \& {Vegter}}]{Feldbrugge2019}
{Feldbrugge}, J., {van Engelen}, M., {van de Weygaert}, R., {Pranav}, P., {Vegter}, G., 2019, \jcap, 2019, 9, 052

\bibitem[{{Fluri} et~al.(2019){Fluri} \& {Kacprzak} et~al.}]{Fluri2019}
{Fluri}, J., {Kacprzak}, T., {Lucchi}, A., et~al., 2019, \prd, 100, 6, 063514

\bibitem[{{Fluri} et~al.(2018){Fluri} \& {Kacprzak} \& {Refregier} \& {Amara} \& {Lucchi} \& {Hofmann}}]{Fluri2018}
{Fluri}, J., {Kacprzak}, T., {Refregier}, A., {Amara}, A., {Lucchi}, A., {Hofmann}, T., 2018, \prd, 98, 12, 123518

\bibitem[{{Foreman-Mackey} et~al.(2013){Foreman-Mackey} \& {Hogg} \& {Lang} \& {Goodman}}]{Foreman-Mackey2013}
{Foreman-Mackey}, D., {Hogg}, D.~W., {Lang}, D., {Goodman}, J., 2013, \pasp, 125, 306

\bibitem[{{Gatti} et~al.(2020){Gatti} \& {Chang} et~al.}]{G20}
{Gatti}, M., {Chang}, C., {Friedrich}, O., et~al., 2020, \mnras, 498, 3, 4060

\bibitem[{{Gatti} et~al.(2022{\natexlab{a}}){Gatti} \& {Giannini} et~al.}]{y3-sourcewz}
{Gatti}, M., {Giannini}, G., {Bernstein}, G.~M., et~al., 2022{\natexlab{a}}, \mnras, 510, 1, 1223

\bibitem[{{Gatti} et~al.(2022{\natexlab{b}}){Gatti} \& {Jain} et~al.}]{moments2021}
{Gatti}, M., {Jain}, B., {Chang}, C., et~al., 2022{\natexlab{b}}, \prd, 106, 8, 083509

\bibitem[{Gatti et~al.(2024)Gatti \& Jeffrey et~al.}]{paperA}
Gatti, M., Jeffrey, N., Whiteway, L., et~al., 2024, Phys. Rev. D, 109, 063534

\bibitem[{{Gatti} et~al.(2024){Gatti} \& {Jeffrey} et~al.}]{sourceclustering}
{Gatti}, M., {Jeffrey}, N., {Whiteway}, L., et~al., 2024, \mnras, 527, 1, L115, \eprint arXiv:{2307.13860}

\bibitem[{{Gatti} et~al.(2021){Gatti} \& {Sheldon} et~al.}]{y3-shapecatalog}
{Gatti}, M., {Sheldon}, E., {Amon}, A., et~al., 2021, \mnras, 504, 3, 4312

\bibitem[{{G{\'o}rski} et~al.(2005){G{\'o}rski} \& {Hivon} et~al.}]{2005ApJ...622..759G}
{G{\'o}rski}, K.~M., {Hivon}, E., {Banday}, A.~J., et~al., 2005, \apj, 622, 2, 759, \eprint arXiv:{astro-ph/0409513}

\bibitem[{{Harnois-D{\'e}raps} et~al.(2022){Harnois-D{\'e}raps} \& {Martinet} \& {Reischke}}]{HD2022}
{Harnois-D{\'e}raps}, J., {Martinet}, N., {Reischke}, R., 2022, \mnras, 509, 3, 3868

\bibitem[{{Heavens} et~al.(2000){Heavens} \& {Jimenez} \& {Lahav}}]{Heavens2000}
{Heavens}, A.~F., {Jimenez}, R., {Lahav}, O., 2000, \mnras, 317, 965

\bibitem[{{Heydenreich} et~al.(2022){Heydenreich} \& {Br{\"u}ck} et~al.}]{Heydenreich2022}
{Heydenreich}, S., {Br{\"u}ck}, B., {Burger}, P., et~al., 2022, \aap, 667, A125

\bibitem[{{Heydenreich} et~al.(2021){Heydenreich} \& {Br{\"u}ck} \& {Harnois-D{\'e}raps}}]{Heydenreich2021}
{Heydenreich}, S., {Br{\"u}ck}, B., {Harnois-D{\'e}raps}, J., 2021, \aap, 648, A74

\bibitem[{{Huff} \& {Mandelbaum}(2017)}]{HuffMcal2017}
{Huff}, E., {Mandelbaum}, R., 2017, arXiv e-prints, 1702.02600

\bibitem[{{Jeffrey} et~al.(2021{\natexlab{a}}){Jeffrey} \& {Alsing} \& {Lanusse}}]{jeffrey_lfi}
{Jeffrey}, N., {Alsing}, J., {Lanusse}, F., 2021{\natexlab{a}}, \mnras, 501, 1, 954

\bibitem[{{Jeffrey} et~al.(2021{\natexlab{b}}){Jeffrey} \& {Gatti} et~al.}]{y3-massmapping}
{Jeffrey}, N., {Gatti}, M., {Chang}, C., et~al., 2021{\natexlab{b}}, \mnras, 505, 3, 4626

\bibitem[{{Jeffrey} \& {Wandelt}(2020)}]{momentnets}
{Jeffrey}, N., {Wandelt}, B.~D., 2020, Third Workshop on Machine Learning and the Physical Sciences, NeurIPS 2020, arXiv:2011.05991

\bibitem[{{Jeffrey} et~al.(2024){Jeffrey} \& {Whiteway} et~al.}]{jeffrey2024dark}
{Jeffrey}, N., {Whiteway}, L., {Gatti}, M., et~al., 2024, arXiv e-prints, arXiv:2403.02314

\bibitem[{{Kacprzak} et~al.(2023){Kacprzak} \& {Fluri} \& {Schneider} \& {Refregier} \& {Stadel}}]{cosmogrid1}
{Kacprzak}, T., {Fluri}, J., {Schneider}, A., {Refregier}, A., {Stadel}, J., 2023, \jcap, 2023, 2, 050

\bibitem[{{Kacprzak} et~al.(2016){Kacprzak} \& {Kirk} et~al.}]{Kacprzak2016}
{Kacprzak}, T., {Kirk}, D., {Friedrich}, O., et~al., 2016, \mnras, 463, 3653

\bibitem[{{Kaiser} \& {Squires}(1993)}]{KaiserSquires}
{Kaiser}, N., {Squires}, G., 1993, \apj, 404, 441

\bibitem[{{Kratochvil} et~al.(2010){Kratochvil} \& {Haiman} \& {May}}]{Kratochvil2010}
{Kratochvil}, J.~M., {Haiman}, Z., {May}, M., 2010, \prd, 81, 4, 043519

\bibitem[{{Kratochvil} et~al.(2012){Kratochvil} \& {Lim} \& {Wang} \& {Haiman} \& {May} \& {Huffenberger}}]{Kratochvil2012}
{Kratochvil}, J.~M., {Lim}, E.~A., {Wang}, S., {Haiman}, Z., {May}, M., {Huffenberger}, K., 2012, \prd, 85, 10, 103513

\bibitem[{{Li} et~al.(2023){Li} \& {Hoekstra} et~al.}]{li2023kids}
{Li}, S.-S., {Hoekstra}, H., {Kuijken}, K., et~al., 2023, \aap, 679, A133, \eprint arXiv:{2306.11124}

\bibitem[{Li et~al.(2023)Li \& Zhang et~al.}]{HSC2}
Li, X., Zhang, T., Sugiyama, S., et~al., 2023, Phys. Rev. D, 108, 123518

\bibitem[{{Liu} et~al.(2015){Liu} \& {Petri} \& {Haiman} \& {Hui} \& {Kratochvil} \& {May}}]{Liu2015}
{Liu}, J., {Petri}, A., {Haiman}, Z., {Hui}, L., {Kratochvil}, J.~M., {May}, M., 2015, \prd, 91, 6, 063507

\bibitem[{{Lu} et~al.(2023){Lu} \& {Haiman} \& {Li}}]{Lu2023}
{Lu}, T., {Haiman}, Z., {Li}, X., 2023, \mnras, 521, 2, 2050

\bibitem[{{MacCrann} et~al.(2022){MacCrann} \& {Becker} et~al.}]{y3-imagesims}
{MacCrann}, N., {Becker}, M.~R., {McCullough}, J., et~al., 2022, \mnras, 509, 3, 3371, \eprint arXiv:{2012.08567}

\bibitem[{Mallat(1999)}]{Mallat1999}
Mallat, S., 1999, A wavelet tour of signal processing, Elsevier

\bibitem[{{Mallat}(2011)}]{Mallat2012}
{Mallat}, S., 2011, arXiv e-prints, arXiv:1101.2286

\bibitem[{{Mallat} et~al.(2020)}]{Mallat2020}
{Mallat}, S., et~al., 2020, Information and Inference: A Journal of the IMA, 9, 3, 721, ISSN 2049-8772

\bibitem[{{Martinet} et~al.(2018){Martinet} \& {Schneider} et~al.}]{Martinet2018}
{Martinet}, N., {Schneider}, P., {Hildebrandt}, H., et~al., 2018, \mnras, 474, 1, 712

\bibitem[{{Muir} et~al.(2020){Muir} \& {Bernstein} et~al.}]{y3-blinding}
{Muir}, J., {Bernstein}, G.~M., {Huterer}, D., et~al., 2020, \mnras, 494, 3, 4454

\bibitem[{{Myles} et~al.(2021){Myles} \& {Alarcon} et~al.}]{y3-sompz}
{Myles}, J., {Alarcon}, A., {Amon}, A., et~al., 2021, \mnras, 505, 3, 4249

\bibitem[{Papamakarios et~al.(2017)Papamakarios \& Pavlakou \& Murray}]{Papamakarios2017}
Papamakarios, G., Pavlakou, T., Murray, I., 2017, Advances in neural information processing systems, 30

\bibitem[{{Parroni} et~al.(2020){Parroni} \& {Cardone} \& {Maoli} \& {Scaramella}}]{Parroni2020}
{Parroni}, C., {Cardone}, V.~F., {Maoli}, R., {Scaramella}, R., 2020, \aap, 633, A71

\bibitem[{{Parroni} et~al.(2021){Parroni} \& {Tollet} \& {Cardone} \& {Maoli} \& {Scaramella}}]{Parroni2021}
{Parroni}, C., {Tollet}, {\'E}., {Cardone}, V.~F., {Maoli}, R., {Scaramella}, R., 2021, \aap, 645, A123

\bibitem[{{Peel} et~al.(2018){Peel} \& {Pettorino} \& {Giocoli} \& {Starck} \& {Baldi}}]{Peel2018}
{Peel}, A., {Pettorino}, V., {Giocoli}, C., {Starck}, J.-L., {Baldi}, M., 2018, \aap, 619, A38

\bibitem[{{Petri} et~al.(2015){Petri} \& {Liu} \& {Haiman} \& {May} \& {Hui} \& {Kratochvil}}]{Petri2015}
{Petri}, A., {Liu}, J., {Haiman}, Z., {May}, M., {Hui}, L., {Kratochvil}, J.~M., 2015, \prd, 91, 10, 103511

\bibitem[{{Planck Collaboration}(2020)}]{aghanim19}
{Planck Collaboration}, 2020, \aap, 641, A5, \eprint arXiv:{1907.12875}

\bibitem[{{Porqueres} et~al.(2022){Porqueres} \& {Heavens} \& {Mortlock} \& {Lavaux}}]{Porqueres2022}
{Porqueres}, N., {Heavens}, A., {Mortlock}, D., {Lavaux}, G., 2022, \mnras, 509, 3, 3194

\bibitem[{{Porth} \& {Smith}(2021)}]{Porth2021}
{Porth}, L., {Smith}, R.~E., 2021, \mnras, 508, 3, 3474

\bibitem[{Potter et~al.(2017)Potter \& Stadel \& Teyssier}]{potter2017pkdgrav3}
Potter, D., Stadel, J., Teyssier, R., 2017, Computational Astrophysics and Cosmology, 4, 1, 2

\bibitem[{Raveri \& Doux(2021)}]{Raveri:2021wfz}
Raveri, M., Doux, C., 2021, Phys. Rev. D, 104, 4, 043504, \eprint arXiv:{2105.03324}

\bibitem[{Raveri \& Hu(2019)}]{Raveri:2018wln}
Raveri, M., Hu, W., 2019, Phys. Rev. D, 99, 4, 043506, \eprint arXiv:{1806.04649}

\bibitem[{{Ribli} et~al.(2019){Ribli} \& {Pataki} \& {Csabai}}]{Ribli2018}
{Ribli}, D., {Pataki}, B.~{\'A}., {Csabai}, I., 2019, Nature Astronomy, 3, 93

\bibitem[{{S{\'a}nchez} et~al.(2022){S{\'a}nchez} \& {Prat} et~al.}]{y3-shearratio}
{S{\'a}nchez}, C., {Prat}, J., {Zacharegkas}, G., et~al., 2022, \prd, 105, 8, 083529, \eprint arXiv:{2105.13542}

\bibitem[{{Schneider} \& {Teyssier}(2015)}]{Schneider2015}
{Schneider}, A., {Teyssier}, R., 2015, \jcap, 2015, 12, 049

\bibitem[{{Scolnic} et~al.(2018){Scolnic} \& {Jones} et~al.}]{scolnic18}
{Scolnic}, D.~M., {Jones}, D.~O., {Rest}, A., et~al., 2018, \apj, 859, 2, 101, \eprint arXiv:{1710.00845}

\bibitem[{{Secco} et~al.(2022){Secco} \& {Samuroff} et~al.}]{y3-cosmicshear2}
{Secco}, L.~F., {Samuroff}, S., {Krause}, E., et~al., 2022, \prd, 105, 2, 023515

\bibitem[{{Shan} et~al.(2018){Shan} \& {Liu} et~al.}]{Shan2018}
{Shan}, H., {Liu}, X., {Hildebrandt}, H., et~al., 2018, \mnras, 474, 1, 1116

\bibitem[{{Sheldon} \& {Huff}(2017)}]{SheldonMcal2017}
{Sheldon}, E.~S., {Huff}, E.~M., 2017, \apj, 841, 24

\bibitem[{{Thiele} et~al.(2020){Thiele} \& {Hill} \& {Smith}}]{Thiele2020}
{Thiele}, L., {Hill}, J.~C., {Smith}, K.~M., 2020, \prd, 102, 12, 123545

\bibitem[{{Valogiannis} \& {Dvorkin}(2022{\natexlab{a}})}]{Valogiannis2022}
{Valogiannis}, G., {Dvorkin}, C., 2022{\natexlab{a}}, \prd, 106, 10, 103509

\bibitem[{{Valogiannis} \& {Dvorkin}(2022{\natexlab{b}})}]{Valogiannis2022a}
{Valogiannis}, G., {Dvorkin}, C., 2022{\natexlab{b}}, \prd, 105, 10, 103534

\bibitem[{{Van Den Berg}(1999)}]{VDB1999}
{Van Den Berg}, J., 1999, Wavelets in Physics, Cambridge University Press

\bibitem[{{Van Waerbeke} et~al.(2013){Van Waerbeke} \& {Benjamin} et~al.}]{VanWaerbeke2013}
{Van Waerbeke}, L., {Benjamin}, J., {Erben}, T., et~al., 2013, \mnras, 433, 3373

\bibitem[{{Vicinanza} et~al.(2016){Vicinanza} \& {Cardone} \& {Maoli} \& {Scaramella} \& {Er}}]{Vicinanza2016}
{Vicinanza}, M., {Cardone}, V.~F., {Maoli}, R., {Scaramella}, R., {Er}, X., 2016, arXiv e-prints, arXiv:1606.03892

\bibitem[{{Vicinanza} et~al.(2018){Vicinanza} \& {Cardone} \& {Maoli} \& {Scaramella} \& {Er}}]{Vicinanza2018}
{Vicinanza}, M., {Cardone}, V.~F., {Maoli}, R., {Scaramella}, R., {Er}, X., 2018, \prd, 97, 2, 023519

\bibitem[{{Vicinanza} et~al.(2019){Vicinanza} \& {Cardone} \& {Maoli} \& {Scaramella} \& {Er} \& {Tereno}}]{Vicinanza2019}
{Vicinanza}, M., {Cardone}, V.~F., {Maoli}, R., {Scaramella}, R., {Er}, X., {Tereno}, I., 2019, \prd, 99, 4, 043534

\bibitem[{{Z{\"u}rcher} et~al.(2023){Z{\"u}rcher} \& {Fluri} \& {Ajani} \& {Fischbacher} \& {Refregier} \& {Kacprzak}}]{Zuercher2022}
{Z{\"u}rcher}, D., {Fluri}, J., {Ajani}, V., {Fischbacher}, S., {Refregier}, A., {Kacprzak}, T., 2023, \mnras, 525, 1, 761, \eprint arXiv:{2206.01450}

\bibitem[{{Z{\"u}rcher} et~al.(2021){Z{\"u}rcher} \& {Fluri} \& {Sgier} \& {Kacprzak} \& {Refregier}}]{Zuercher2021}
{Z{\"u}rcher}, D., {Fluri}, J., {Sgier}, R., {Kacprzak}, T., {Refregier}, A., 2021, \jcap, 2021, 1, 028

\bibitem[{{Z{\"u}rcher} et~al.(2022){Z{\"u}rcher} \& {Fluri} et~al.}]{Zuercher2021b}
{Z{\"u}rcher}, D., {Fluri}, J., {Sgier}, R., et~al., 2022, \mnras, 511, 2, 2075

\end{thebibliography}
\bibliographystyle{mn2e_2author_arxiv_amp.bst}

%%%%%%%%%%%%%%%%%%%%%%%%%%%%%%%%%%%%%%%%%%%%%%%%%%

%%%%%%%%%%%%%%%%% APPENDICES %%%%%%%%%%%%%%%%%%%%%

\appendix

\section{$\Lambda$CDM validation}\label{ref:validation_LCDM}

In this Appendix we perform a number of validation tests for the $\Lambda$CDM analysis, similar to what was done in Paper A for the $w$CDM analysis. In particular, we:

\begin{itemize}
\item conduct a baryonic contamination test, i.e. we verify that the scales used in the $\Lambda$CDM analysis are not affected by potential unmodelled baryonic feedback processes;
\item test that additive biases due to PSF modelling errors were negligible at the data vector level;
\item check that potential errors in our model for the source clustering effects were negligible;
\item perform a coverage test to ensure that the size of the posteriors we recover is accurate;
\item carry out an end-to-end $\Lambda$CDM analysis on an independent set of simulations (CosmogridV1 sims), demonstrating that we recover the true cosmology of the simulations.
\end{itemize}

Beyond these tests, we also add two sets of tests where:
\begin{itemize}
    \item we show we can recover the true cosmology of the simulations even if the parameters describing redshift uncertainties and shear bias calibrations were $2\sigma$ off their mean values;
    \item we show that our posteriors are not sensitive to the details of the simulations used (box size, number of particles, redshift resolution).
\end{itemize}

The initial test we conducted was the baryonic contamination test. This followed the same methodology as presented in Paper A, albeit tailored for a $\Lambda$CDM analysis. Our analysis does not include a model for the effects of baryons at small scales. Consequently, we need to ensure that the scales used are not compromised by baryonic effects. To this end, we employed a suite of simulations (\texttt{CosmoGridV1}) that have been post-processed to include baryonic feedback using the baryonic correction model \citep{Schneider2015,Arico2020}. Each simulation is available in two versions: one with baryonification and one without. We generate multiple instances of the DES maps using both versions to calculate our summary statistics.  We then test whether the posteriors of cosmological parameters derived from data vectors including baryonic feedback do not exhibit significant bias when compared to those from data vectors without baryonic effects. We follow the criterion adopted by the DES Y3 cosmological analysis (\citealt{y3-cosmicshear1}; \citealt*{y3-cosmicshear2}; \citealt{y3-3x2ptkp}), which requires that the peak of the marginalized two-dimensional posterior of $\Omega_{\textrm{m}}$ and $S_8$ from the analysis of baryon-contaminated data must fall within $0.3\sigma$ of that from clean data.  Table~\ref{fig:contamination} reports the contamination levels obtained using all available scales of our analyses, for all the individual summary statistics and certain combinations of them. Notably, all the summary statistics stayed within our pre-established limits for contamination, thereby showing the robustness of our analysis against potential baryonic feedback effects.

For the tests addressing potential PSF contamination, we emulated the methodology from \paperA{} and analyzed two sets of mock data from the \texttt{CosmoGridV1} suite, i.e. with and without added contributions from PSF modelling errors as estimated in \cite{y3-shapecatalog}. We confirmed that, even when focusing on a $\Lambda$CDM analysis, none of our summary statistic combinations showed a bias in the $S_8 \text{--} \Omega_{\textrm{m}}$ plane larger than 0.10$\sigma$. This indicates that PSF modelling errors are insignificant for the scale range considered in our study.
 
Regarding the tests for potential errors in modelling source clustering, we analyzed two sets of maps with assumed galaxy-matter biases of $b=0.5$ and $b=1.5$, diverging from the standard value of unity used in our analysis. We found that for all our summary statistics combinations, the bias in the $S_8 \text{--} \Omega_{\textrm{m}}$ plane was under 0.10$\sigma$. Thus, the influence on cosmological parameters is minimal, confirming that our source clustering model is accurate enough for our analysis.

\begin{figure}
	\includegraphics[width=\columnwidth]{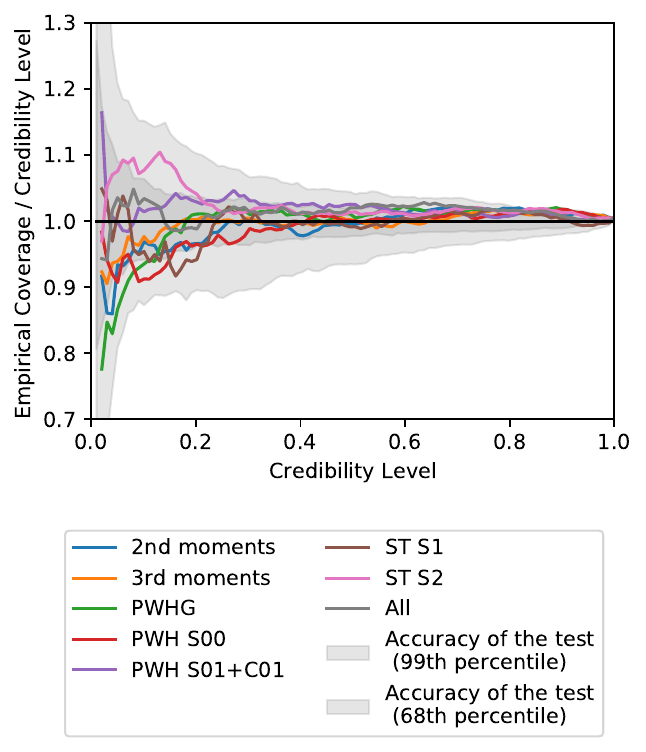}
    \caption{Expected coverage probability of posteriors derived from the simulation-based inference pipeline, using various summary statistics, with respect to the credibility level. The grey shaded areas represent the precision of the test, constrained by the finite number of posteriors used in this analysis.} %\LW{In the figure label, should `accuracy' be `precision'?}}
    \label{fig:empirical}
\end{figure}

The next test aimed at ensuring that the confidence levels from simulation-based inference were correctly estimated. This involved carrying out an empirical coverage test. In such tests, the inference process is repeated multiple times to confirm that the estimated posteriors accurately reflect the true parameter probabilities. For our purposes, we ran the inference multiple (100) times, each time omitting one mock data vector (chosen at random) from the neural likelihood estimation. We specifically omitted mocks with $w=-1$, corresponding to the $\Lambda$CDM scenario, from the \texttt{DarkGridV1} simulation set. The likelihood for the excluded data vector was then calculated, the posterior determined, and its accuracy verified against the known parameter values. We assessed coverage probabilities within the three-dimensional parameter space of $\Omega_{\rm m}$, $S_8$, $A_{\textrm{IA}}$ using the \textsc{tarp} package, which applies the `Tests of Accuracy with Random Points' (TARP) method for estimating the coverage probabilities of generative posterior estimators. The outcomes, illustrated in Fig.~\ref{fig:empirical}, reveal that the expected coverage corresponds with the credibility levels at 5 percent, signifying that our posterior estimates are properly calibrated.

\begin{figure}
	\includegraphics[width=\columnwidth]{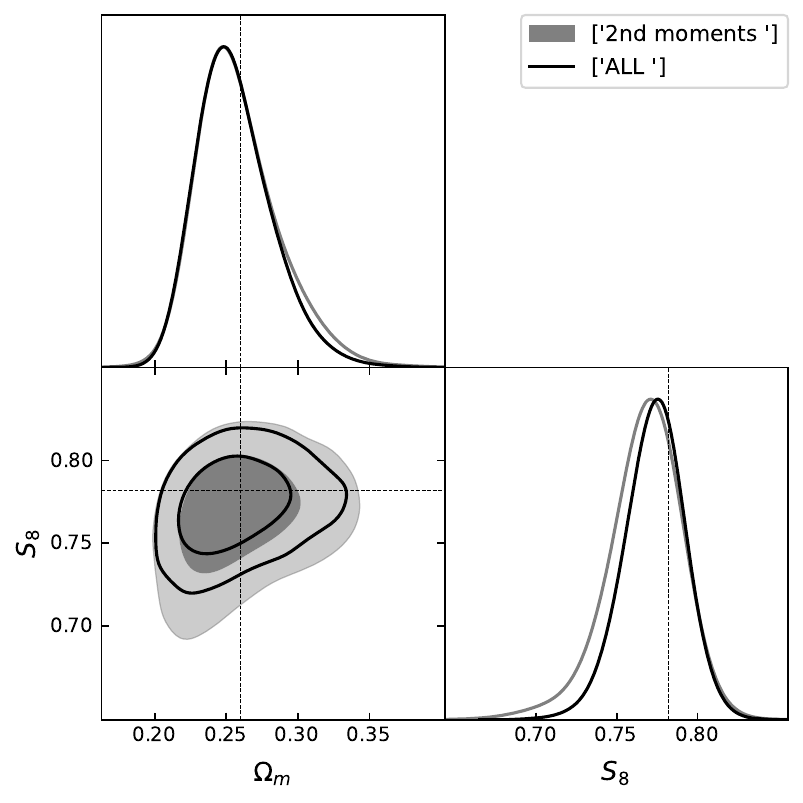}
    \caption{Posterior distributions of the cosmological parameters $\Omega_{\textrm{m}}$ and $S_8$ for different second moments and for the combinations of all the summary statistics considered in this work, as measured in \texttt{CosmoGridV1} simulations. The dotted black lines indicate the values of the cosmological parameters in the simulations. The two-dimensional marginalised contours in these figures show the 68 percent and 95 percent credible regions.}
    \label{fig:end_to_end}
\end{figure}

We then confirm our pipeline's ability to accurately recover the actual cosmology from a set of simulations not previously used in its construction. For this purpose, we use 400 independent DES Y3 mock catalogues, generated by the \texttt{CosmoGridV1} simulations, identical to those employed for the $w$CDM analysis validation in \paperA{}. None of these \texttt{CosmoGridV1} simulations are used during the training of our inference pipeline (i.e. during the compression or when training the NDEs), but they are rather used as a `target' data vector. Each mock has the same cosmology; we further assume no intrinsic alignment, while for the other nuisance parameters (shear calibration and redshift uncertainties) we assume values at the centre of the priors. By measuring all summary statistics across the mocks and taking their average, we mitigate the influence of noise. The resulting posterior distributions for the cosmological parameters $\Omega_{\textrm{m}}$ and $S_8$, derived from the second moments and the combination of all summary statistics, are depicted in Fig.~\ref{fig:end_to_end}. The simulated analysis recovers the true cosmological parameters of the simulations, underscoring the reliability of our analytical framework. 

\begin{figure}
	\includegraphics[width=\columnwidth]{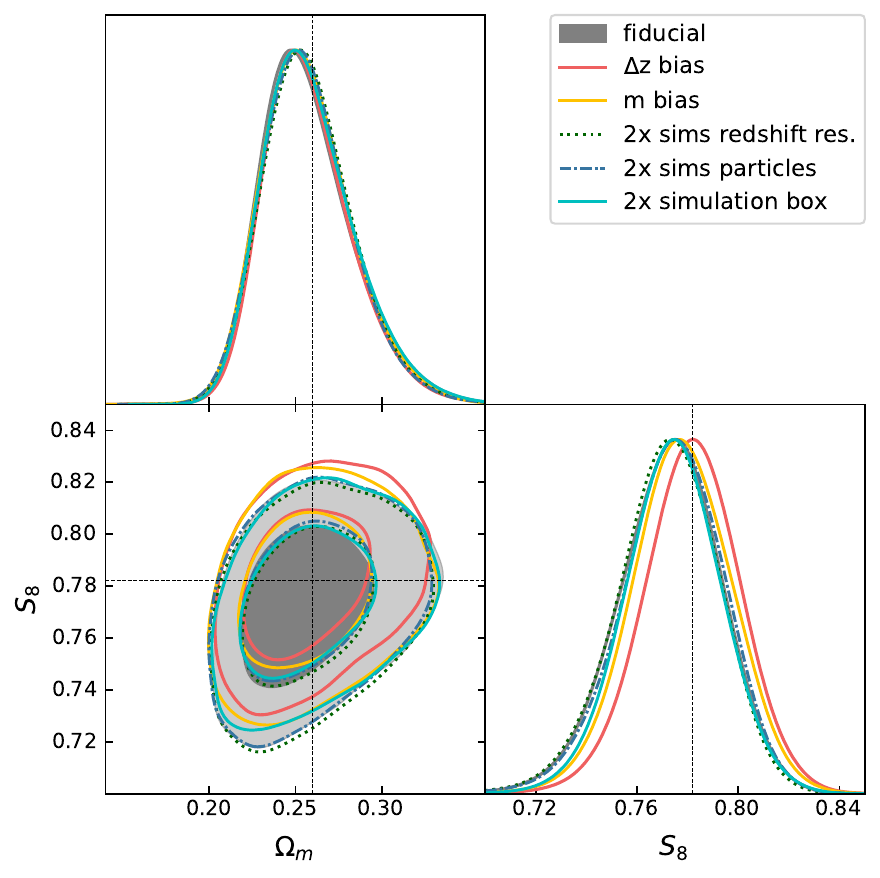}
    \caption{Posterior distributions of the cosmological parameters $\Omega_{\textrm{m}}$ and $S_8$ for different combinations of all the summary statistics considered in this work, as measured in a number of different \texttt{CosmoGridV1} simulations.}
    \label{fig:end_to_end_tests}
\end{figure}

In the final series of tests, we evaluate our ability to accurately recover the cosmological parameters of simulations under various conditions. Specifically, we assess whether the correct cosmology can be identified when: 1) the simulation includes shear biases that deviate by $2\sigma$ from the average values of their prior distributions; 2) the redshift mean values diverge by $2\sigma$ from their prior averages; 3) the simulations use double the number of particles; 4) the simulations have been produced with double the box size; and 5) the simulations feature twice the redshift resolution. For these latter three scenarios, we used benchmark simulations from the \texttt{CosmoGridV1} suite. The results, depicted in Fig. \ref{fig:end_to_end_tests}, confirm that in every instance we successfully retrieved the true cosmological values of the simulations.

\begin{table}
%\tiny
\caption{Bias in the parameter posteriors assessed by contrasting the results from an analysis that simulated mocks with baryonic feedback against one using mocks with no such feedback. The impact on various summary statistics are quantified by measuring the separation between the peaks of the posterior distributions in the $S_8 \text{--}\Omega_{\text{m}}$ plane. We found that all biases are well below the $0.3\sigma$ threshold (the upper limit of bias tolerated in 
our analysis).}
\centering
%\begin{adjustbox}{width=0.7\textwidth}
\begin{tabular}{l c}
 \toprule
 \textbf{Summary Statistic(s)} & \textbf{Contamination}\\
 & \textbf{$S_8-\Omega_{\textrm{m}}$}\\
 \midrule
2nd moments & 0.06$\sigma$ \\
$\WPHG$  & 0.05$\sigma$ \\
3rd moments & 0.05$\sigma$ \\
WPH S00 & 0.01$\sigma$ \\
WPH S01+C01 & 0.05$\sigma$ \\
WPH S00+S01+C01 & 0.11$\sigma$ \\
ST1 & 0.05$\sigma$ \\
ST2 & 0.03$\sigma$ \\
ST1+ST2 & 0.04$\sigma$ \\
2nd+3rd moments & 0.10$\sigma$ \\
2nd moments+WPH & 0.05$\sigma$ \\
2nd moments+ST1 & 0.08$\sigma$ \\
2nd+3rd moments+ST+WPH & 0.15$\sigma$ \\
\bottomrule
\end{tabular}
%\end{adjustbox}
\label{fig:contamination}
\end{table}

\section{B-modes test}\label{ref:B_modes}

\begin{figure}
\begin{center}
\includegraphics[width=0.45 \textwidth]{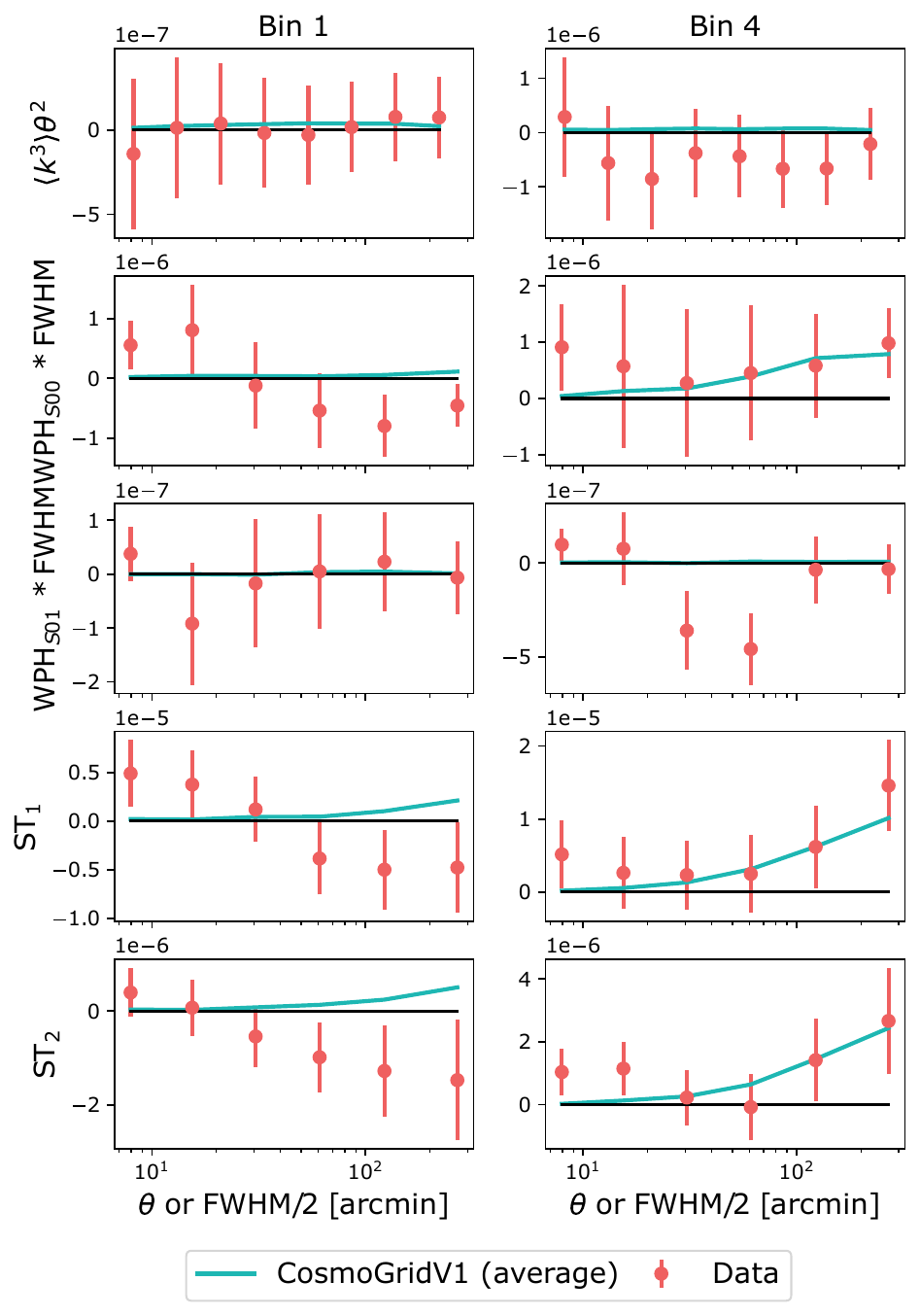}
\end{center}
\caption{Comparison of the measured summary statistics of B-mode convergence maps in our data with those obtained from the \texttt{CosmoGridV1} simulations. We did not include measurements for second moments and WPHG in this comparison, as these statistics are already incorporated into the data vectors used for the cosmological analysis.}
\label{fig:B-modes}
\end{figure}

At first order, weak lensing is not expected to produce B-modes. Consequently, B-modes serve as a null test for identifying potential systematic effects not accounted for in the analysis. However, due to our map-making procedure using the Kaiser-Squires algorithm, B-modes can emerge as a result of masking effects, as noted \cite{moments2021}. This occurs because a small fraction of the E-mode power leaks into the B-mode map, predominantly at large scales. In our data vector, we included the second moments and WHPG of the B-mode maps, although their contribution to the cosmological constraints is small. We opted not to include other summary statistics for B-modes, as this would double the length of the data vector, complicating the data compression without significantly improving our constraints. Nevertheless, we conducted a comparison of B-modes in our data against those measured in simulations for the summary statistics beyond second moments and WHPG. Such a comparison is illustrated in Figure \ref{fig:B-modes}. Specifically, we compared against measurements from the \texttt{CosmoGridV1} simulations, which were performed at a fixed cosmology. The measurements from our data and the simulations show good agreement. We did not make a significant detection of B-modes for third moments and WHP S01. These results on third moments corroborate the findings of \cite{moments2021}. For other statistics (ST1, ST2, and WPH S00), both data and simulations indicate a slight, non-null B-mode signal at large scales, likely attributable to E-mode leakage caused by masking effects. Given the strong correspondence between simulations and data, we conclude that our B-mode analysis reveals no evidence of unaccounted systematics in our simulations.

\section{NDEs and parameters posterior}\label{sect:indivuals}

\begin{figure*}
\includegraphics[width=0.45\textwidth]{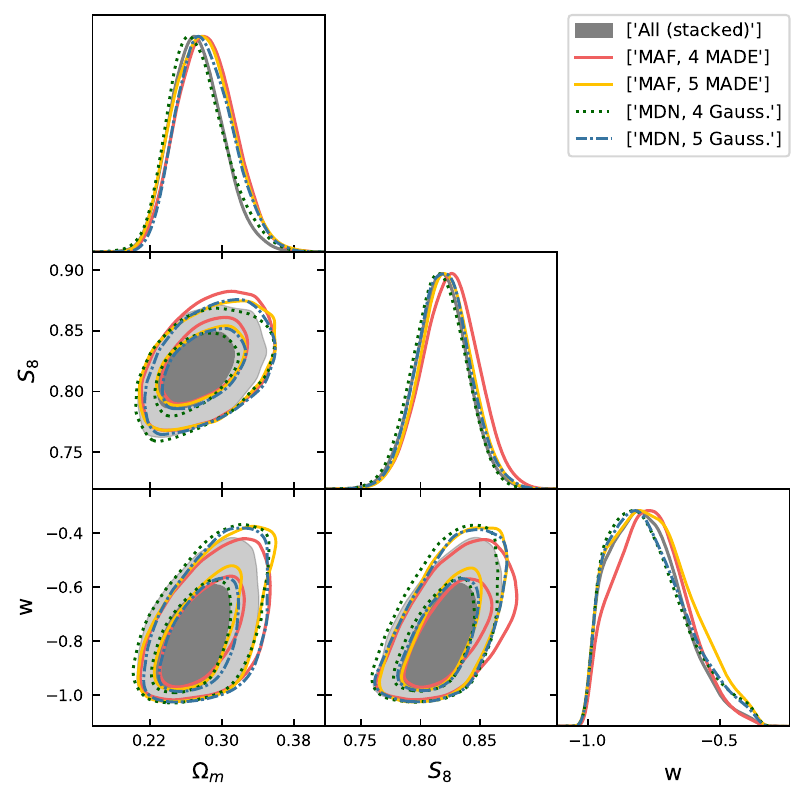}
\includegraphics[width=0.45\textwidth]{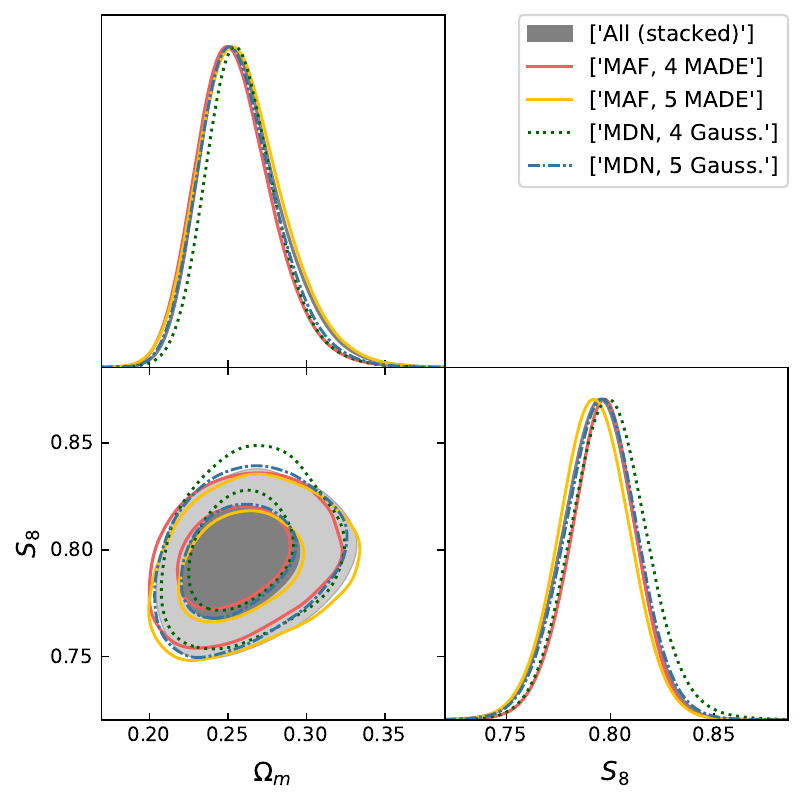}
\caption{Posterior distributions of the cosmological parameters $\Omega_{\textrm{m}}$ and $\sigma_8$ for the combination of all the summary statistics considered in this work, as measured in data, for the $\Lambda$CDM (right) and $w$CDM (left) analyses. We show the different posteriors as estimated by the different NDEs used in this work; we also show their stacked combination (the fiducial setup used in the other Figures of this paper). The dotted black lines indicate the values of the cosmological parameters in the simulations. The two-dimensional marginalised contours in these figures show the 68 percent and 95 percent credible regions.}\label{fig:individual_posterior}
\end{figure*}

In this work we used four different neural density estimators (NDEs) to estimate the posteriors. In particular, we used two different Gaussian Mixture Density Networks (MDNs) and two different Masked Autoencoders for Distribution Estimation (MADEs). Whenever we showed a posterior or reported the constraints on some parameters in this work, we always obtained these by stacking the four different NDEs. Assuming all the NDEs are flexible enough to describe our likelihood surface, they should all agree in the limit in which the number of simulations used for training becomes large. Fig.~\ref{fig:individual_posterior} shows the posteriors obtained by each individual NDE for our most constraining case (i.e. the combination of all summary statistics); the posteriors are very similar, indicating that our posteriors estimates are robust and stable. 

\section{Impact of the \texttt{DarkGridV1} simulation suite }\label{Appendix_dark_gower}

\begin{figure}
	\includegraphics[width=\columnwidth]{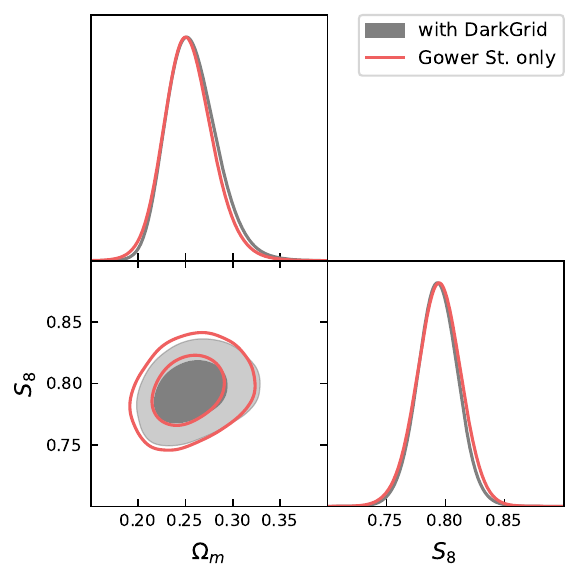}
    \caption{Posterior distributions of the cosmological parameters $\Omega_{\textrm{m}}$ and $S_8$ obtained including or not \texttt{DarkGridV1} simulations into our analysis setup. The posteriors are shown for the $\Lambda$CDM scenario, and for the case where all the summary statistics are combined together.}
    \label{fig:darkgrid_no}
\end{figure}

Our analysis uses simulations in two distinct ways: for training the compression algorithm and for learning the likelihood surface with the NDEs. In this work, we have expanded the simulation suite used in our validation study (\paperA{}), by including simulations from the \texttt{DarkGridV1} suite. These additional simulations are specific to the $\Lambda$CDM model, meaning they do not vary the parameter $w$. Furthermore, they do not vary other parameters such as $h_{100}$, $n_s{\textrm{m}}$, $\Omega_{\textrm{b}}$, or the neutrino mass (unlike those in the Gower Street suite). Our NDEs do not explicitly learn the dependence on these latter parameters; instead, these parameters are effectively marginalized over according to the distribution followed by the simulations. Incorporating the \texttt{DarkGridV1} suite simulations thus modifies the distribution of mocks, which could impact the posterior. Figure \ref{fig:darkgrid_no} shows a comparison of our posterior on the data for the $\Lambda$CDM case, with and without including the \texttt{DarkGridV1} suite in learning the likelihood surface. The posteriors are very similar, indicating that the impact of the differing parameter distribution is negligible.

\section{Cosmological constraints of individual statistics}\label{Appendix_individual} 

\begin{figure*}
\includegraphics[width=0.4\textwidth]{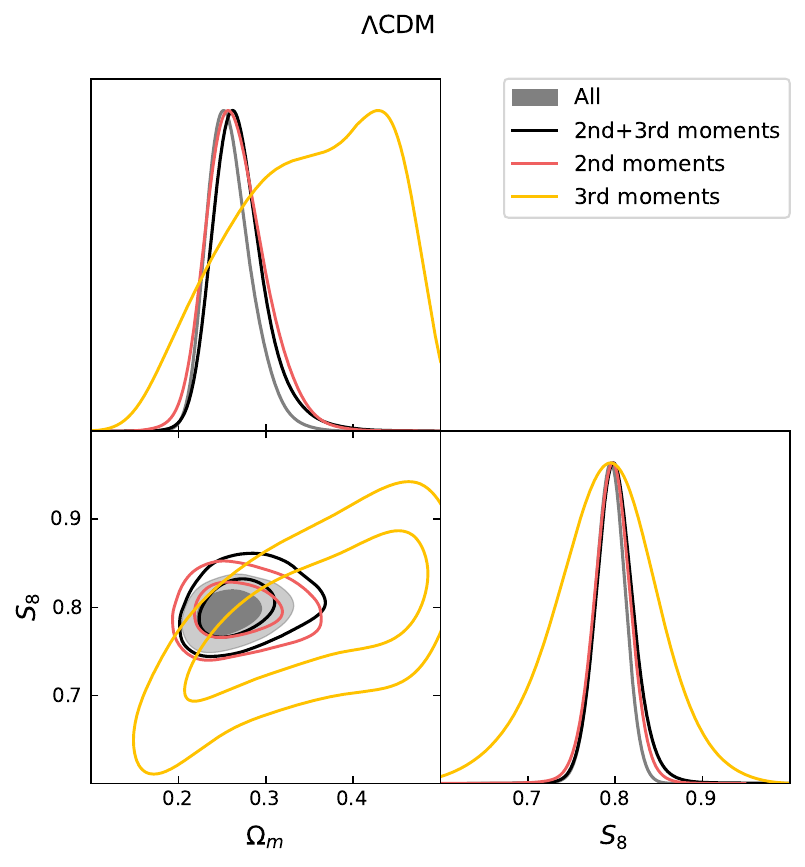}
\includegraphics[width=0.4\textwidth]{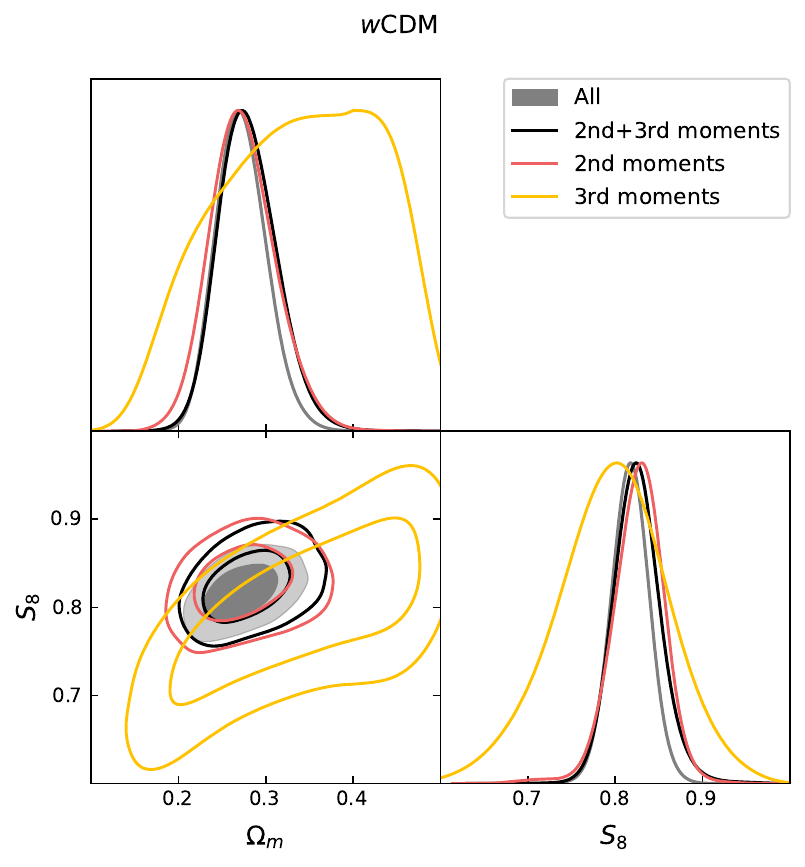}
 \includegraphics[width=0.4\textwidth]{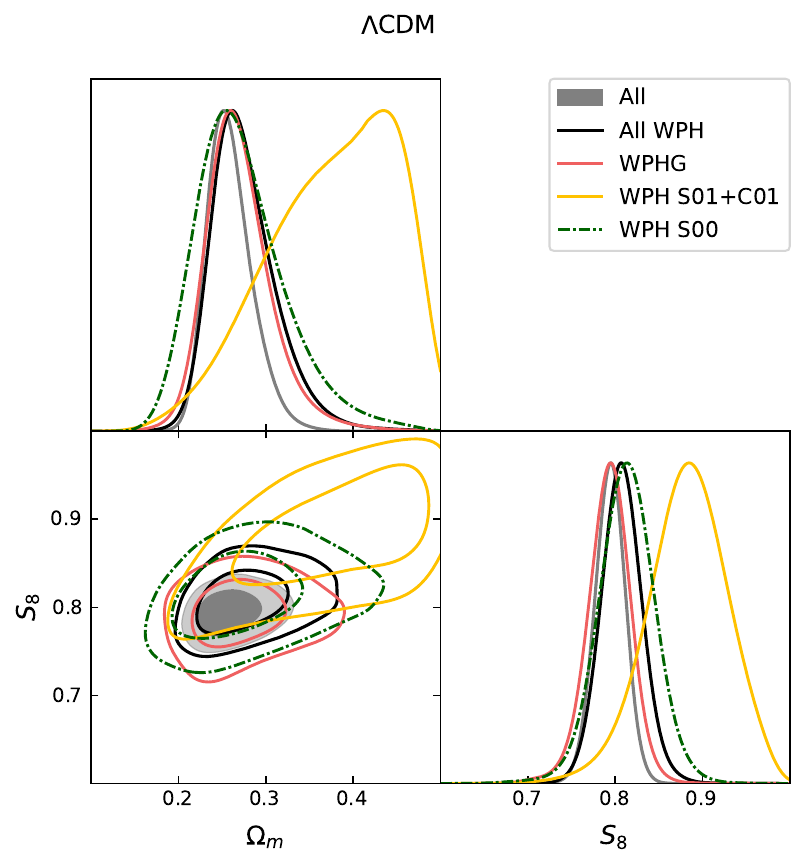}
 \includegraphics[width=0.4\textwidth]{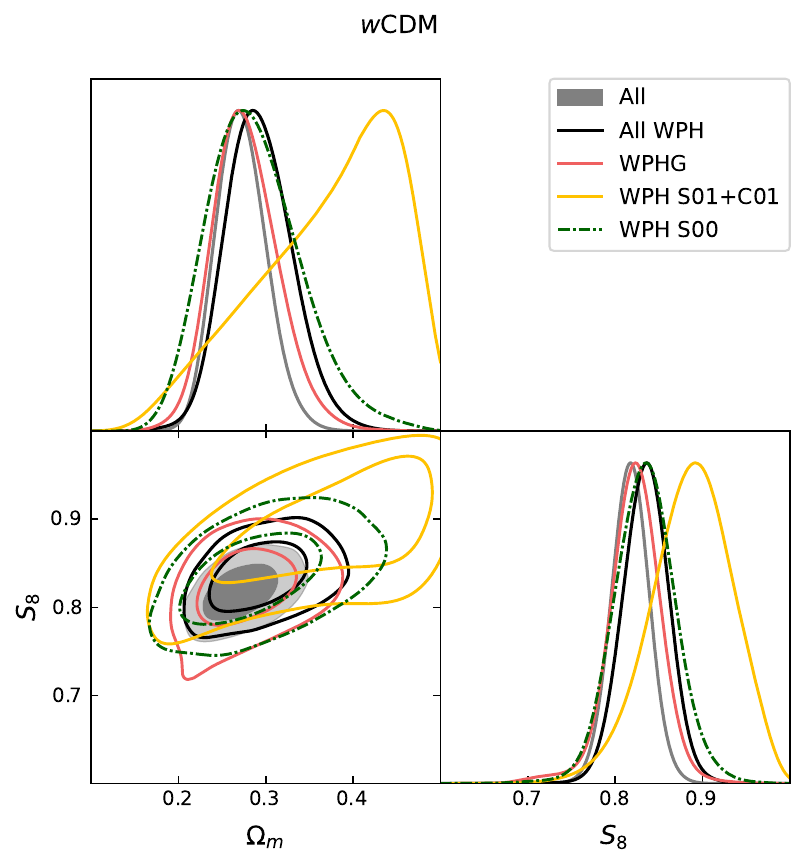}
 \includegraphics[width=0.4\textwidth]{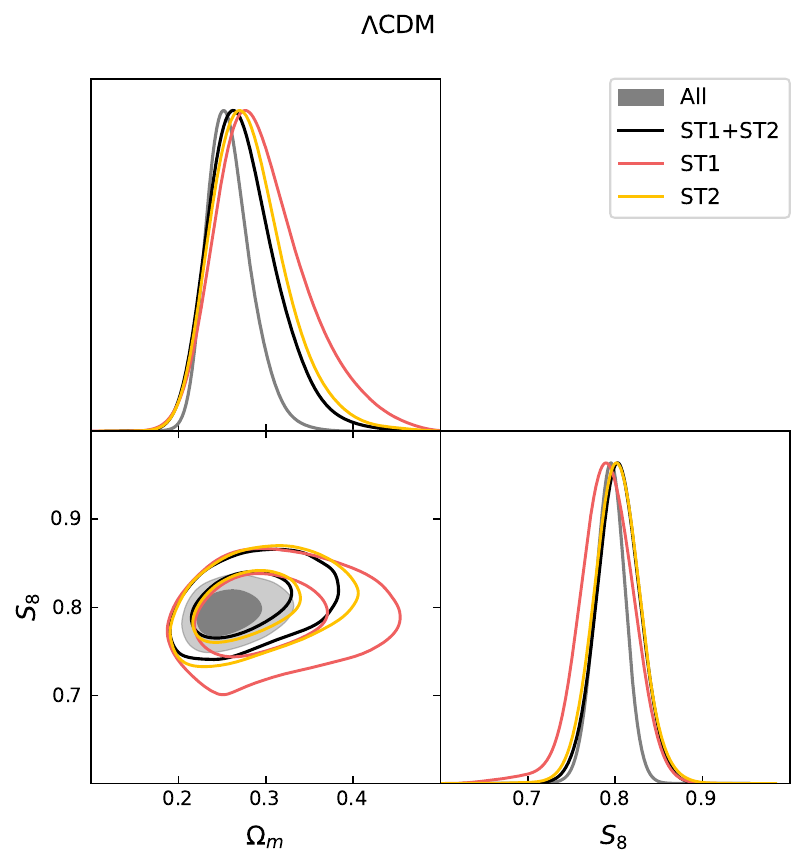}
 \includegraphics[width=0.4\textwidth]{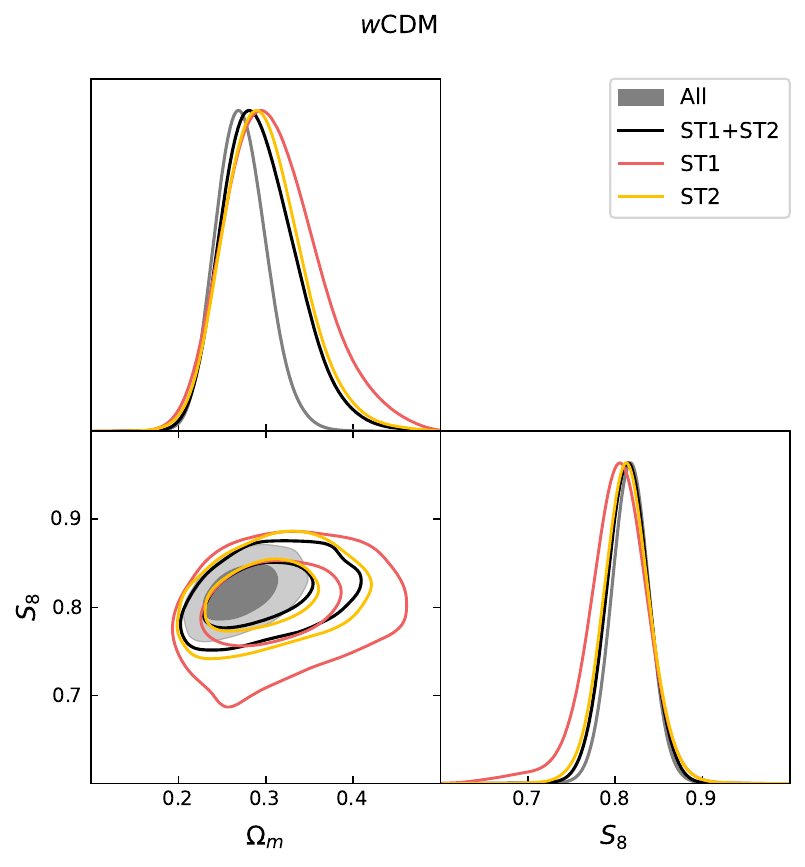}
    \caption{Posterior distributions of the cosmological parameters $\Omega_{\textrm{m}}$ and $S_8$ from individual statistics and their combination, for both $w$CDM (right) and $\Lambda$CDM (left).}
    \label{fig:_}
\end{figure*}

We show in this Appendix the posteriors for $\Omega_{\textrm{m}}$ and $S_8$ obtained from individual statistics, i.e. second moments, third moments, WPHG, WPH S01, WPH S00, ST1, and ST2, and their combination. The posteriors are shown in Fig. \ref{fig:_}, for both $w$CDM (left) and $\Lambda$CDM (right). Constraints from individual probes are largely overlapping and consistent with each other. Third moments and WPH S01 are the less constraining statistic among all those explored here; this is partially because all the other non-Gaussian statistics (ST1, ST2 and WPH S00) also probe part of the Gaussian information of the field, and hence are more constraining.

%\appendix
%\input{Appendix}

%\input{convergence_t}
%\section*{Affiliations}
%\input{aff.tex}

%%%%%%%%%%%%%%%%%%%%%%%%%%%%%%%%%%%%%%%%%%%%%%%%%%

%\begin{equation}
%\avg{\textbf{e_{\textrm{obs}}}} \sim \textbf{\gamma}
%\end{equation}
% Don't change these lines
% \bsp	% typesetting comment
\label{lastpage}
\end{document}